\documentclass[%
 reprint,
 amsmath,amssymb,
 aps,
 floatfix
]{revtex4-2}

\usepackage{graphicx}
\usepackage{dcolumn}
\usepackage{bm}
\usepackage{orcidlink}
\usepackage{siunitx}
\usepackage{xcolor}

\begin{document}

\preprint{arXiv:2507.18955}

\title{A New Framework to Detect Multi-Messenger Signals from Bright Sporadic Stochastic Gravitational Wave Background}

\author{Hardik Jitendra Kuralkar\orcidlink{0009-0002-1868-9507}}
\email{kuralkarhardik@gmail.com}
\affiliation{Department of Physics, Indian Institute of Science Education and Research (IISER) Bhopal,\\
Bhopal Bypass Road, Bhauri, Bhopal - 462066, Madhya Pradesh, India}

\author{Mohit Raj Sah\orcidlink{0009-0005-9881-1788}}
\email{mohit.sah@tifr.res.in}
\author{Suvodip Mukherjee\orcidlink{0000-0002-3373-5236}}
\email{suvodip@tifr.res.in}
\affiliation{Department of Astronomy \& Astrophysics, Tata Institute of Fundamental Research,\\
1, Dr. Homi Bhabha Road, Mumbai - 400005, Maharashtra, India}

\date{\today}

\begin{abstract}
The temporal dependence of the astrophysical stochastic gravitational-wave (GW) background (SGWB) in the hecto-hertz band brings a unique avenue to identify multi-messenger signals to these sources by using coincident detection in both GW and multi-band EM signals. We developed a new analysis pipeline, \textit{Multi-messenger Cross-Correlation} (MC$^2$) that can search for EM counterparts to the SGWB signal originating from both modeled and unmodeled sources by harnessing the nearly full-sky gamma-ray sky map. We provide an observation strategy that can be followed by current and future missions to discover EM counterparts to the weak GW signal hidden in the SGWB. We demonstrate the ability of this technique to drastically reduce the false alarm rates when involving EM multi-band analysis. This formalism aims towards advancing the multi-messenger observation frontier and improving our understanding of the population of bright SGWB sources present in the high-redshift universe and can also be applied to other messengers such as neutrinos in the future.
\end{abstract}

\maketitle


\section{Introduction} \label{sec:intro}

The direct detection of gravitational waves (GWs) by the LIGO-Virgo-KAGRA (LVK) collaboration \cite{abbott_observation_2016} paved the way for exploring a wide range of physics, ranging from astrophysics to fundamental physics. {Since the first detection, the LVK has observed over 300 compact binary coalescences (CBCs), including binary black holes (BBHs), binary neutron stars (BNSs), and neutron star-black hole (NSBH) mergers till the end of O4b \cite{LIGOScientific:2026sit}.} On August 17, 2017, the LVK detector network made its first detection of a BNS merger \cite{abbott_gw170817:_2017}. This landmark discovery was accompanied by the detection of a short gamma-ray burst (sGRB), GRB170817, by Fermi-GBM \citep{Meegan_2009,Goldstein_2017} and INTEGRAL \citep{Savchenko_2017} from the same sky position, around two seconds after the event. Subsequent targeted searches revealed counterparts across multiple lower-frequency electromagnetic (EM) bands, including X-rays \cite{Troja_2017, Fong_2017_GCN, Troja_2017_GCNe, Haggard_2017_GCNb, Margutti_2017, Haggard_2017}, UV/optical/IR \cite{Soares-Santos_2017,Metzger_2017}, and radio \cite{Alexander_2017_GCN,Mooley_2017_GCN,Corsi_2017_GCN,Hallinan_2017}. This marked the first joint detection and analysis of GWs and their EM counterparts, thus initiating the era of multi-messenger astrophysics using GWs. In addition to BNS mergers, EM emissions can arise from other types of CBCs, detectable by LIGO-Virgo \cite{TheLIGOScientific:2014jea,Martynov:2016fzi,Tse_2019,Acernese_2014,Acernese_2019} and upcoming LISA \cite{LISA_2017}, such as NSBH mergers or interactions between compact binaries and their surrounding environments \cite{Li:1998bw,Metzger_Berger_2012,Barbieri:2019sjc,2041-8205-752-1-L15,McKernan:2019hqs,Graham_2020,Mukherjee:2019oma,Veres:2019hsd,Blandford_1977,Zhang_2016,Haiman:2018brf,Palenzuela:2010nf,Farris:2014zjo,Gold:2014dta,Armitage:2002uu}.

GWs are not only detected as individual signals from CBCs but are also expected to contribute to a cumulative signal known as the SGWB. This astrophysical background arises from the sporadic contribution of numerous faint GW events. Among these are faint CBC signals, some of which, such as BNSs and NSBHs, may have well-detected EM counterparts yet remain buried within the SGWB noise, particularly in those EM bands in which sources can be detected beyond redshift $z=0.3$ (which is the typical horizon to detect bright GW events from current generation detectors \cite{KAGRA:2013rdx}). Additionally, the SGWB may include contributions from unmodeled or unknown astrophysical sources. Detecting EM counterparts associated with such weak GW events can offer valuable insights into a variety of high-energy astrophysical processes. To address this challenge, we propose the idea of identifying GW signals embedded in noise by performing a cross-correlation between the SGWB and EM observations at a particular sky direction across multiple wavelength bands in the time domain.

In this paper, we present the technique of multi-messenger time-domain cross-correlation and the computational implementation of this technique using a Python-based software $\mathbf{MC^2}$ (\textbf{M}ulti-messenger \textbf{C}ross-\textbf{C}orrelation). This technique performs time-domain cross-correlation, dubbed TDCC hereafter, between the GW signal and EM signal from the transient sources to identify the potential EM counterparts of events hidden in the SGWB. This formalism is also extended to unmodeled/unknown potential GW sources that are not yet detected, which are beyond the current detection limits of LVK. The technique can also mitigate glitches that mark their persistent presence in the LIGO detectors. This idea opens a new avenue in multi-messenger observation science. We also supplement this idea with an observation strategy that can be followed by current and future multi-messenger missions to advance the detection frontier. 

The structure of the paper is as follows: In Section \ref{sec:motiv}, we discuss the motivation behind the formalism. Section \ref{sec:obs_strategy} discusses the observation and follow-up strategies for GW signals. In Section \ref{sec:formalism}, we give a mathematical overview of the formalism. We provide an insight into the event properties and modeling of signals ranging from GW signals, their EM counterparts, and high-energy radiations in Section \ref{sec:modelling}, while Section \ref{sec:application} contains our results on application to simulated data. An account of false alarm rates (FAR) and their mitigation is given in Section \ref{sec:FAR detect}. {Estimation for expected detection rate of BNS events using TDCC is highlighted in \ref{sec:estimated-event-rate}.} We conclude the paper with a brief summary in Section \ref{sec:Conclusion}.

\section{Motivation} \label{sec:motiv}

Only a fraction of compact binary mergers occurring in the universe are detectable by our ground-based gravitational-wave detectors \cite{KAGRA:2013rdx}. LVK detectors are only capable of detecting signals from loud and nearby events, which is a very small fraction of the events happening in the universe. With aLIGO design sensitivity, we can only detect individual BNS events up to about redshift $z \sim 0.1$ \cite{KAGRA:2013rdx}. Therefore, BNS and NSBH sources beyond this redshift constitute the sub-threshold GW signal and the SGWB signal. On the other hand, several ground-based and space-based telescopes are continuously monitoring the EM transient sky in all spectra and high-energy particles (like cosmic rays and neutrinos). Some bands, gamma-ray and radio emissions in particular, using, for example, Fermi-GBM \cite{fermi} and VLA \cite{VLA}, respectively, have detection capabilities beyond $z=0.1$. As a result, while many GW events themselves remain undetected by current GW detectors, their EM counterparts \cite{Bogdanovi_2022,Nakar_2020}, if they have any, may be detected by EM telescopes. Such shrouded GW events can be identified by exploiting multi-band correlations between EM and GW signals.

If the SGWB were composed of a single, isolated BNS signal, the optimal detection strategy would be a coherent matched-filtering search. In such a scenario, matched filtering exploits the full phase coherence of the waveform and is known to provide the highest signal-to-noise ratio for compact binary coalescence. However, we aim to address situations where the GW signal is very weak, sub-threshold, or embedded in a background composed of multiple unresolved events, making direct coherent detection challenging or infeasible. In these cases, cross-correlation techniques between GW and EM signals become valuable. Our pipeline $\mathbf{MC^2}$ is designed to exploit multi-messenger coincidences, even when individual GW signals are too weak to be detected through matched filtering.

The detection of the EM counterpart can enable precise localization of the GW source, including measurement of its redshift and identification of the host galaxy. Therefore, the synergy between the GW and EM probes is very crucial to reinforce the study of the CBCs that are known to have an EM counterpart, such as BNS, NSBH, and BBH mergers in AGN disc.

\section{Observation Strategies} \label{sec:obs_strategy}
\begin{figure}
    \centering
    \includegraphics[width=\linewidth]{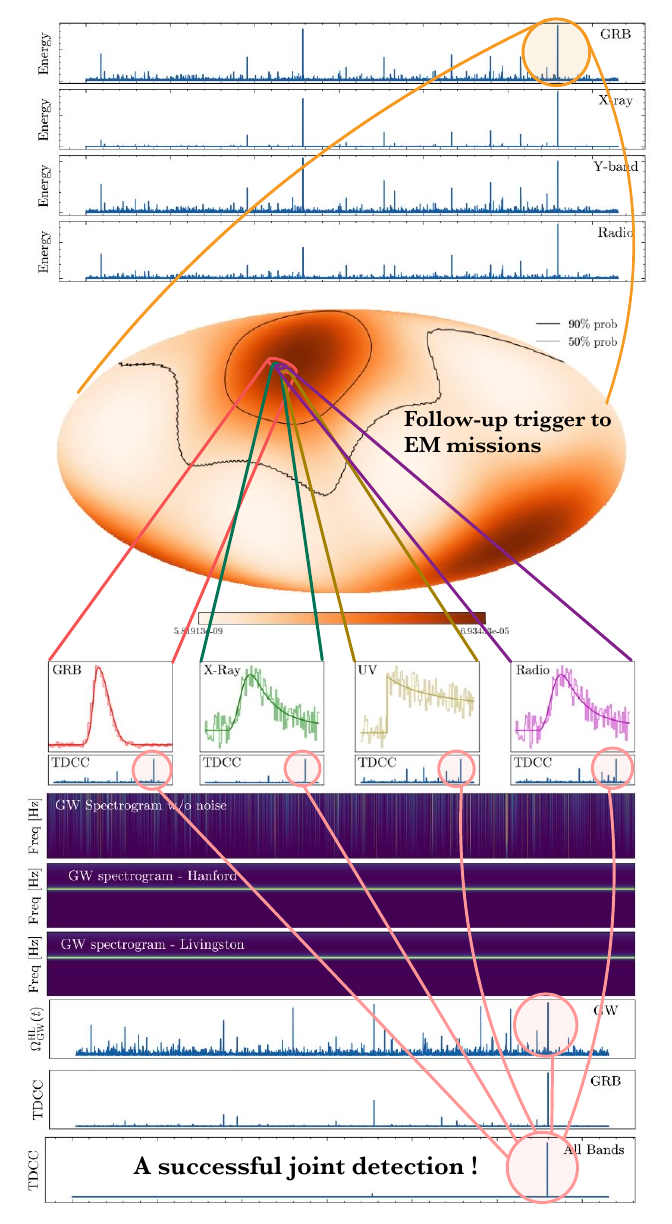}
    \caption{Flowchart showing the $\mathbf{MC^2}$ technique to detect the GW events from the GW background using the cross-correlation of the background signal with the multi-messenger EM signal.}
    \label{fig:flowchart}
\end{figure}

This section outlines the observation strategies for future multi-messenger detection using the proposed $\mathbf{MC^2}$ framework. GRBs are among the most luminous and promptly detectable EM signals associated with compact object mergers, making them valuable triggers for multi-messenger follow-up. Leveraging this feature, we propose the following observation strategy to identify potential GW counterparts to gamma-ray signals and, subsequently, to other EM bands:
\begin{itemize}
    \item \textbf{Prompt GRB Detection and EM Follow-up}:
    Upon the detection of a GRB, a follow-up trigger is issued to other EM observatories to search for coincident emissions in complementary bands, such as X-ray, optical, or radio, either in real time or during late-time afterglow phases. This coordinated observational effort ensures broad spectral coverage and increases the likelihood of identifying EM signatures across different emission timescales and energy ranges.
    \item \textbf{SGWB Estimation from Noisy Strain Data}:
    Simultaneously, the SGWB energy density, $\Omega_{\text{GW}}(f,t)$, can be estimated by cross-correlating the noisy strain data from two or more GW detectors (e.g., Hanford and Livingston). This cross-correlation technique enhances sensitivity to weak and unmodeled GW signals by reducing uncorrelated instrumental noise and revealing coherent astrophysical contributions.
    \item \textbf{TDCC Between SGWB and Gamma-ray Signals}:
    A TDCC is then performed between the reconstructed $\Omega_{\text{GW}}(f,t)$ and the gamma-ray time series. If a significant correlation is observed within a physically plausible time-delay window, it may indicate a GW counterpart to the GRB event. This step serves as the primary filter for identifying candidate multi-messenger events from otherwise noisy GW backgrounds. 
    \item \textbf{Extension of TDCC to Other EM Bands}:
    Once a correlation between the SGWB and the GRB is established, the same TDCC formalism is extended to other EM channels. Correlating the SGWB signal with X-ray, optical, and radio observations, ideally from the same sky location and time window, further strengthens the confidence in the multi-messenger association and reduces the probability of false alarms. This layered approach increases the robustness of detection by incorporating temporal, spectral, and spatial coincidences across a broad range of observatories.
\end{itemize}

The GW signal of an event, which is heavily shrouded in background noise, can be extracted by cross-correlating the background signals between two detectors (more on this in Appendix \ref{apx:sgwb}). This strategy is illustrated in Figure \ref{fig:flowchart}. The first four panels in the figure present the energy versus time profiles for different EM bands: GRBs, X-rays, Y-band (optical), and radio. A specific transient event is highlighted in the GRB panel, which serves as the reference point for multi-wavelength analysis. For this GRB event, we search for temporally coincident emissions in other EM bands originating from the same sky location, as indicated in the corresponding skymap. The emission waveforms for the X-ray, optical, and radio bands are shown alongside their respective TDCC plots, with the GW signal in the panels directly below the skymap.

Three spectrograms follow, showing the GW strain measurements from the Hanford and Livingston detectors, both with and without added noise, to illustrate the observational conditions. The simulations for GW are performed as per Section \ref{sec:popmodels}, and we assume the A+ sensitivity of the detector \citep{barsotti2018a+}. Below these, we present the time-dependent GW energy density, $\Omega_{\text{GW}}(t)$, which serves as the input for the TDCC analysis. Upon establishing a statistically significant correlation between the GW signal and the GRB emission, we proceed to perform TDCC with the remaining EM bands. The final panel highlights a prominent correlation signature, marking a successful joint detection and the identification of a potential multi-messenger event involving both GW and EM counterparts.

In this framework, EM observations play a crucial role in guiding the search for corresponding GW signals within the large GW data volume. EM data are used as triggers because EM telescopes can probe significantly higher redshifts than current GW detectors. The proposed TDCC across multiple EM frequency bands further helps discriminate true associations from unrelated transient sources. The TDCC signal becomes nonzero only when a common signal is present in both GW and EM data. While unrelated transients typically appear in a single EM band, a true associated event produces correlated signatures across multiple EM frequency bands.

Furthermore, TDCC between gamma-ray and GW data can identify mutually associated signals, enabling discrimination between unrelated transients and those linked to GW sources, while reducing sky localization from several hundred to a few tens of square degrees. This improved localization allows targeted follow-up searches over smaller sky regions using lower-frequency EM channels such as X-ray, UV, optical, infrared, and radio, to identify sources exhibiting time-domain correlation with subthreshold GW and gamma-ray signals. Moreover, the proposed method enables the extraction of weak GW signals from noise using EM information, which is not currently achievable. The success of this technique relies on the availability of multi-band, high-cadence EM observations covering a large sky fraction and extending to high cosmological redshifts, as well as on networks of GW detectors.

\section{Formalism}  \label{sec:formalism}

The possibility of the temporal correlations between the multi-frequency EM signal and the faint GW signal submerged in the background was first explored by \cite{mukherjee-fundamental-2021}. In this section, we discuss the TDCC formalism between GW and EM signals. The TDCC can be performed directly between the cross-correlated time series data between two GW detectors and the EM flux.
\begin{widetext}
    \begin{equation}
        C_{\nu}(t_{\text{obs}}, \Delta t_{\nu}, \hat{\alpha})  = \frac{1}{\delta t} \int_{t_{\text{obs}} - \delta t /2}^{t_{\text{obs}} + \delta t /2} \mathrm{d} t'  \left[ d_{I}(t', \hat{\alpha}) d_{J}(t', \hat{\alpha}) \right] \left[\mathcal{\hat{I}}_{\nu} (t' + \Delta t_{\nu}, \hat{\alpha}) - \mathcal{\overline{I}}_{\nu} \right],
        \label{eqn:TDCC_ht}
    \end{equation}
\end{widetext}
where $d_{I}(t', \hat{\alpha})$ and $d_{J}(t', \hat{\alpha})$ are the time series GW strain data of detectors I and J at time $t'$ coming from the sky direction $\hat{\alpha}$. $\mathcal{\hat{I}}(\nu,t'+\Delta t_{\nu},\hat{\alpha})$ represents the flux of the EM signal at time $t = t'+\Delta t_{\nu}$, and $\mathcal{\overline{I}}_{\nu}$ is the sky- and time-averaged EM signal, $\Delta t_{\nu}$ is the time delay between the arrival of the GW signal and the EM signal at fixed sky direction $\hat{\alpha}$, and $\delta t$ is the signal averaging time. We can also perform the TDCC on the Fourier transform of the GW data, $\tilde{d}_I (f,t)$, where the `$t$' represents the central time of the time bin over which the Fourier transform is performed. We can cross-correlate $\tilde{d}_I (f,t)$ and $\tilde{d}_J (f,t)$ from detectors I and J, respectively, to obtain SGWB density given by
\begin{equation}
    {\begin{aligned}      
        \hat{\Omega}_{\text{GW}}(f,t, \hat{\alpha}) \equiv & ~  \langle \tilde{d}_I (f,t,\hat{\alpha}) \tilde{d}_J^* (f,t,\hat{\alpha}) \rangle,
    \end{aligned}}
\end{equation}
\begin{equation}
    \tilde{d}(f,t) = \int_{t-\delta t/2}^{t+\delta t/2} d(t') e^{-i2 \pi f t'} \mathrm{d}t'.
\end{equation}

We define for the time-domain correlation between the SGWB ($\hat{\Omega}_{\text{GW}}(f,t,\hat{\alpha})$), averaged over the time interval ($\delta t$) centered at time $t$, and the EM signal as
\begin{widetext}
\begin{equation} 
    C_{\nu}(t, \Delta t_{\nu},\hat{\alpha}) = \Delta f \sum\limits_{f} \left( \hat{\Omega}_{\text{GW}} (f, t, \hat{\alpha}) - \overline{\Omega}_{\text{GW}} (f) \right) \frac{1}{{\delta t} }\int\limits_{t-\delta t/2}^{t+\delta t/2} \left( \mathcal{\hat{I}}_{\nu} (t' + \Delta t_{\nu},\hat{\alpha}) - \mathcal{\overline{I}}_{\nu} \right ) \mathrm{d} t',
    \label{eqn:TDCC_Omega}
\end{equation}
\end{widetext}
where $\hat{\Omega}_{\text{GW}}(f,t,\hat{\alpha})$ is the observed SGWB signal, $\overline{\Omega}_{\rm GW}(f)$ is the time-averaged SGWB, and $\Delta f$ is the frequency bin. The TDCC can not only be applied to the SGWB from intermittent signals from compact binary mergers but also to the intermittent burst signals from unknown sources. The application of the TDCC on these signals has been demonstrated in Section \ref{sec:application}. 

{We note that the data entering Eq. \eqref{eqn:TDCC_ht} are whitened, such that the noise power spectral density is approximately flat across frequency. In this case, summing $\hat{\Omega}_{\rm GW}(f, t, \hat{\alpha})$ over frequency bins provides a simple but optimal estimator. The present formulation is therefore best suited to regimes where the detector noise is well-characterized and whitening is effective. In more general situations, where a particular template is followed for GW signals, the estimator can be extended by introducing a frequency-dependent filter $R(f)$, leading to a weighted sum over frequency bins. Wiener filter is one such filter that can be put in this formalism, potentially improving sensitivity beyond the simple unweighted case.
}

The EM counterparts of all these GW events are expected to arrive with a delay relative to the GW signal, and this delay time, $\Delta t_{\nu}$, from a fixed sky direction $\hat{\alpha}$, is anticipated to vary depending on the frequency of the EM signal. For the prompt gamma-ray signal, the value of $\Delta t_{\nu}$ can be less than a second \citep{ligo_scientific_collaboration_multi-messenger_2017,kasliwal_illuminating_2017,troja_afterglow_2019}, whereas for lower frequency EM signals, operating in X-ray, UV, optical, infrared, and radio, the value of $\Delta t_{\nu}$ can be of the order of a few hours to days \citep{kasliwal_illuminating_2017,chase_kilonova_2022}. By using Eqs. \eqref{eqn:TDCC_ht} and \eqref{eqn:TDCC_Omega}, we can search for potential correlations between the SGWB and EM signals. The details on the implementation of this technique to determine the time delay and the averaging time are given in Appendix \ref{sec:implementation}.

Traditionally, EM follow-up searches are triggered by confidently detected GW events and disseminated to multiple telescopes. In contrast, TDCC between transient EM and GW signals yields a significant correlation only when a common signal is present in both datasets; otherwise, the correlation remains consistent with zero. Consequently, EM signals detected across multiple frequency bands and accompanied by a corresponding GW signature can be distinguished from GW detector noise or unrelated transients. If a binary source emitting GWs also produces EM emission at different frequencies $\nu_i$ with time delays $\Delta t_{\nu_i}$, these signals will originate from the same sky location.

Any transient EM signal is captured as an excess in the intensity $\hat{\mathcal{I}}_\nu(t, \hat{\alpha})$ above the background. For multiple EM frequency channels, sky maps of time-dependent intensity variations can be constructed and cross-correlated with the SGWB signal. Performing TDCC between GW and EM data across multiple frequencies is essential to distinguish unrelated transients from genuine EM counterparts. This approach enables the isolation of weak signals from noise and unrelated foregrounds, since noise in different detectors is uncorrelated. Consequently, this provides a robust method for identifying EM counterparts of SGWB sources, with EM observations serving as an effective guide to separate true signals from noise.

\section{Modeling the Multi-messenger signals} \label{sec:modelling}

GW170817 is the only multi-messenger GW event detected so far \cite{abbott_gw170817:_2017}. However, with improved detector sensitivity and the EM counterpart detection strategy, we expect to detect the counterparts more frequently in the future. In this section, we describe the modeling of multi-messenger signals with a brief discussion on the event properties to understand the emission spectra. The details of population models for GWs used for the generation of SGWB data are described in Section \ref{sec:popmodels}. Section \ref{subsec:gw170817} discusses the modeling of EM counterparts to GW events. 

\subsection{Gravitational Waves Modeling} \label{sec:popmodels}
We discuss the SGWB from CBCs and unmodeled GW sources in Section \ref{sec:SGWB_CBC} and Section \ref{sec:gw burst}, respectively.

\subsubsection{SGWB from BNS, NS-BH, BBH} \label{sec:SGWB_CBC}

The only direct evidence of GWs has come from compact binary coalescences, including BBH, BNS, and NSBH mergers \citep{abbott2023gwtc}. The SGWB resulting from compact binary coalescences arises from the superposition of the GW signals from all the merger events of all source properties and formation channels. The SGWB from a population of compact binary mergers is given by \citep{allen1996stochastic,phinney2001practical,christensen2018stochastic}
\begin{widetext}
{\begin{equation}
    \begin{aligned}
     \overline{\Omega}_{\rm{GW}}(f) & = \int \mathrm{d} m_1 P_{1}(m_1)  \int \mathrm{d} m_2 P_{2}(m_2)  \int\limits_{z_{\rm min}}^{\infty} \mathrm{d} z~ \bigg[\frac{\mathrm{d} V}{\mathrm{d} z} \frac{R(z)}{(1+z)}\bigg]  \bigg[\frac{1}{\rho_c c^2}  \frac{(1+z) f_r}{4 \pi d_{\ell}^{2} c} \frac{\mathrm{d} E_{\rm{GW}}(z,m_1,m_2)}{\mathrm{d} f_r} \bigg],
    \end{aligned}
    \label{SGWB}
\end{equation}}
\end{widetext}
where $\rho_c c^2$ is the critical energy density required to close the universe, $d_{\ell}$ represents the luminosity distance of the source, $R(z)$ is the merger rate of the compact binary per unit comoving volume, $P_{1}(m_1)$ and $P_{2}(m_2)$ are the mass distributions of the component masses of the compact binary, and $\frac{\mathrm{d} E_{\rm{GW}}(z,m_1,m_2)}{\mathrm{d} f_r} $ is the energy emitted by the source per unit source frame frequency, {$f_r = f(1+z)$}.

The SGWB is expected to exhibit temporal fluctuations when averaged over short time intervals \citep{Mukherjee:2019oma,Sah_2023}. This behavior arises from the \textit{time-dependent} nature of the signal, where individual merger events contribute intermittently rather than continuously, leading to periods where the signals do not overlap. We define the SGWB averaged over short time intervals ($\delta t$) centered at the time $t$ as
\begin{equation}
    \Omega_{\rm{GW}}(f,t)  = \sum\limits_i  \frac{1}{\delta t} \bigg[\frac{1}{\rho_c c^2}  \frac{(1+z^i) f}{4 \pi d_{\ell}^{2}(z^i) c} \frac{\mathrm{d} E_{\rm{GW}}(z^i,m_1^i,m_2^i)}{\mathrm{d} f} \bigg],
    \label{SGWB2}
\end{equation}
where $\delta t$ is the observation time, and the summation is for all the events in the interval $\delta t$. We simulate the SGWB signal by simulating individual compact binary mergers occurring across all redshifts within the time interval $\delta t$ and then summing the contribution of GWs from all the events \citep{Sah_2023}. Our approach involves calculating the mass distribution and merger rate as functions of redshift. We partition the redshifts into multiple bins, extending to a very high redshift. Using Poisson sampling, we determined the number of events within each bin for a given observation period ($\Delta t$) for a Poissonian distribution with mean $\lambda =  \frac{\mathrm{d} V}{\mathrm{d} z} \frac{R_{\rm{GW}}(z)}{(1+z)}~ \Delta t \Delta z $. For each event, we sample component masses of the binary from our mass distributions $P_{1}(m_1)$ and $P_{2}(m_2)$. Finally, we sum the GW density contributed by all events within the given observation period.

We assume a power-law form of the merger rate and the BH mass distribution given by
\begin{equation} \label{eq:merger-rate-z}
    R(z)  = R(0)~ (1+z)^{\alpha},
\end{equation}
\begin{equation}
    P_{\text{BH}}(m) \propto m^{\beta}.
\end{equation}
However, we take a uniform mass distribution for the neutron stars
\begin{equation}
    P_{\text{NS}}(m) = \text{constant.}
\end{equation}

For our simulations, we assume $\alpha= 2.7 $, $\beta= -2.5 $, and local merger rate $R(0)$ = 100 \si{Gpc^{-3}yr^{-1}} and 50 \si{Gpc^{-3}yr^{-1}}  for BNS and NSBH, respectively. The minimum and maximum masses of the neutron star and BHs are taken as (1.2 $M_{\odot}$, 2.5 $M_{\odot}$) and (5 $M_{\odot}$, 40 $M_{\odot}$), respectively.

\subsubsection{SGWB from unmodeled sources} \label{sec:gw burst}
All the GW events detected so far are from CBCs with a major contribution from BBHs. A number of astrophysical sources may produce burst-like signals, which are short-duration pulses of GWs. These transients can include supernova explosions or yet-unknown sources. Neutron stars or pulsar glitches are also expected to generate burst signals.

GW waveforms generated by CBCs have been modeled using numerical relativity \citep{blackman_numerical_2017,dietrich_high-resolution_2018,arun_higher-order_2009,buonanno_comparison_2009,blanchet_gravitational_2014,mishra_ready--use_2016}. However, for other sources it is not well modeled. We attempt to include such potential sources in our analysis, considering a small subset of waveform models as follows:
\begin{enumerate}
    \item \textbf{Sine-Gaussian}: 
    \begin{equation}
        h_{\text{SG}}(t) = A e^{-(t-t_0)^2 / \tau^2} \sin{(2 \pi f_0 (t-t_0) + \phi)},
    \end{equation}
    which is a Gaussian enveloped sine wave, characterized by a central frequency $f_0$, amplitude $A$, time of arrival or peak-time $t_0$, duration $\tau$, and phase $\phi$ which is uniformly sampled between [0, $2\pi$].
    \item \textbf{Unmodeled Signal}: 
    \begin{equation}
        h_{\text{UM}}(t) = A ~ \sum\limits_{f} \sin{(2 \pi ft)},
    \end{equation}
    which is a superposition of sine waves.
    \item \textbf{Ringdown Like}:
    \begin{equation}
        h_{\text{RD}}(t) = A e^{-(t-t_0) / \tau} \sin{(2 \pi f_0 (t-t_0) + \phi)},
    \end{equation}
    with frequency $f_0$, amplitude $A$, time of arrival or peak-time $t_0$, duration $\tau$, and phase $\phi$ which is uniformly sampled between [0, $2\pi$].
    \item \textbf{Gaussian Pulse}:
    \begin{equation}
        h_{\text{GP}}(t) = A e^{-(t-t_0)^2 / \tau^2},
    \end{equation}
    with amplitude $A$,  time of arrival or peak-time, $t_0$, and duration $\tau$.
\end{enumerate}
The waveforms for sine-Gaussian, ringdown, and Gaussian pulse are adopted from \cite{Powell_2015}. The amplitude of these signals is kept at the order of BNS strain ($\sim 10^{-24}$) to be detectable by LIGO current limits. Our goal is to demonstrate that this formalism can encompass any known or unknown signal. 

\subsection{Modeling the EM Counterparts} \label{subsec:gw170817}
On August 17, 2017, the LIGO-Virgo detector network observed a GW signal from the inspiral of two low-mass compact objects consistent with a BNS merger \citep{abbott_gw170817:_2017}. This GW signal is the loudest yet observed, with a combined signal-to-noise ratio (SNR) of 32.4 and a false-alarm-rate estimate of less than one per $8 \times 10^4$ years. Component masses of the binary were inferred to be between 0.86 and 2.26 $M_{\odot}$ \citep{abbott_gw170817:_2017}, in agreement with masses of known neutron stars.

The detection of GW170817 triggered a campaign of EM follow-up observations, leading to the identiﬁcation of NGC4993 as the host galaxy of GW170817/GRB170817A. The source was localized to a region of the sky 28 $\text{deg}^2$ at a luminosity distance of ($42.9 \pm 3.2$) Mpc, which is consistent with the distance of $40^{+8}_{-14}$ Mpc determined with GW data alone \citep{abbott_gw170817:_2017,abbott_gw170104:_2017,abbott_gw151226:_2016}. A subsequent EM follow-up campaign across the spectrum in the X-ray, ultraviolet, optical, infrared, and radio bands, over hours, days, and weeks, revealed the counterparts. We base our modeling on the obtained counterparts. We outline counterparts in the gamma, X-ray, UV/optical/IR, and radio bands in the appendix \ref{sec:EMmodel}.

\section{Application on Mock} \label{sec:application}

\begin{figure*}[t]
    \centering
    \includegraphics[width=\linewidth]{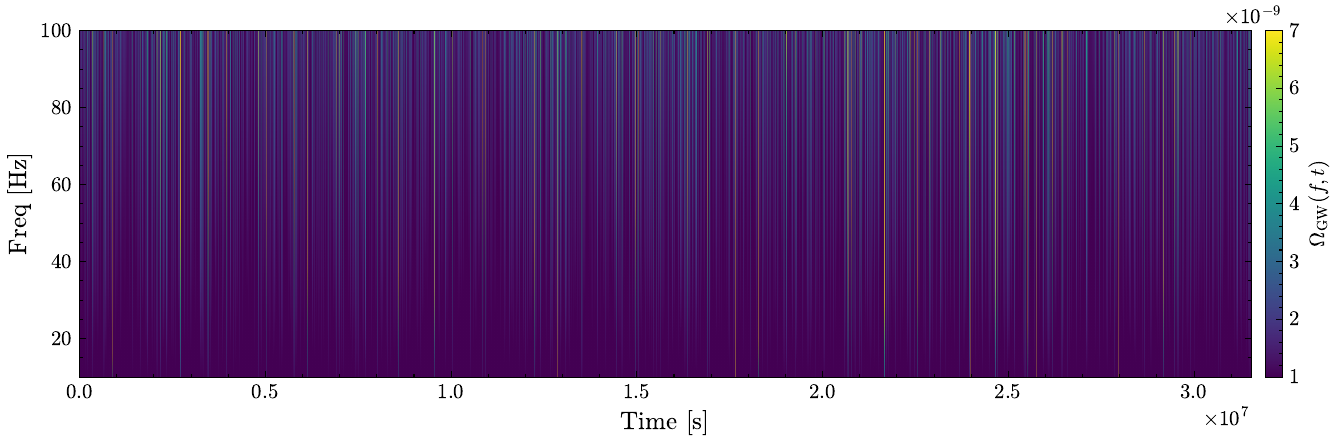}
    \caption{Spectrogram for gravitational wave data simulated for a span of one year averaged over 200s. The bright regions denote higher cumulative GW energy density from the binary mergers in that particular time bin, whereas, the darker regions signify low background.}
    \label{figp:gw_spectrogram}
\end{figure*}

\subsection{TDCC with GW from coalescing binaries}

The SGWB, as discussed earlier, should contain the signals from binary mergers that cannot be individually detected with high matched filtering SNR as an event. The SGWB signal can be obtained by cross-correlating the signal from two different detectors. In this work, we simulate the SGWB signal as per Section \ref{sec:popmodels} with varying signal integration timescale $\Delta t$ and for different merger rates. We assume the A+ sensitivity of the detector \citep{barsotti2018a+}.

The spectrogram of the $\Omega_{\rm GW}(f)$ signal averaged over 200 secs, without noise, for a period of one year is shown in Figure \ref{figp:gw_spectrogram}. The bright regions denote higher GW energy density from the binary mergers in that time bin, whereas the darker regions signify no GW source or low GW energy density.

The TDCC is evaluated by integrating the EM signals over the time bin $\delta t$, corresponding to the signal-averaging interval of the SGWB. A genuine correlation between GW and EM activity manifests as a localized enhancement in the TDCC coefficient, while random fluctuations or noise artifacts may also produce a spurious signal. Such false alarms are more likely in heavily contaminated signals, as they tend to mimic correlated structures in the cross-correlation output. To mitigate this, we cross-correlate the SGWB signal with multiple EM bands simultaneously, as detailed in Section \ref{sec:FAR detect}.

\begin{figure*}[]
    \centering
    \includegraphics[width=\textwidth]{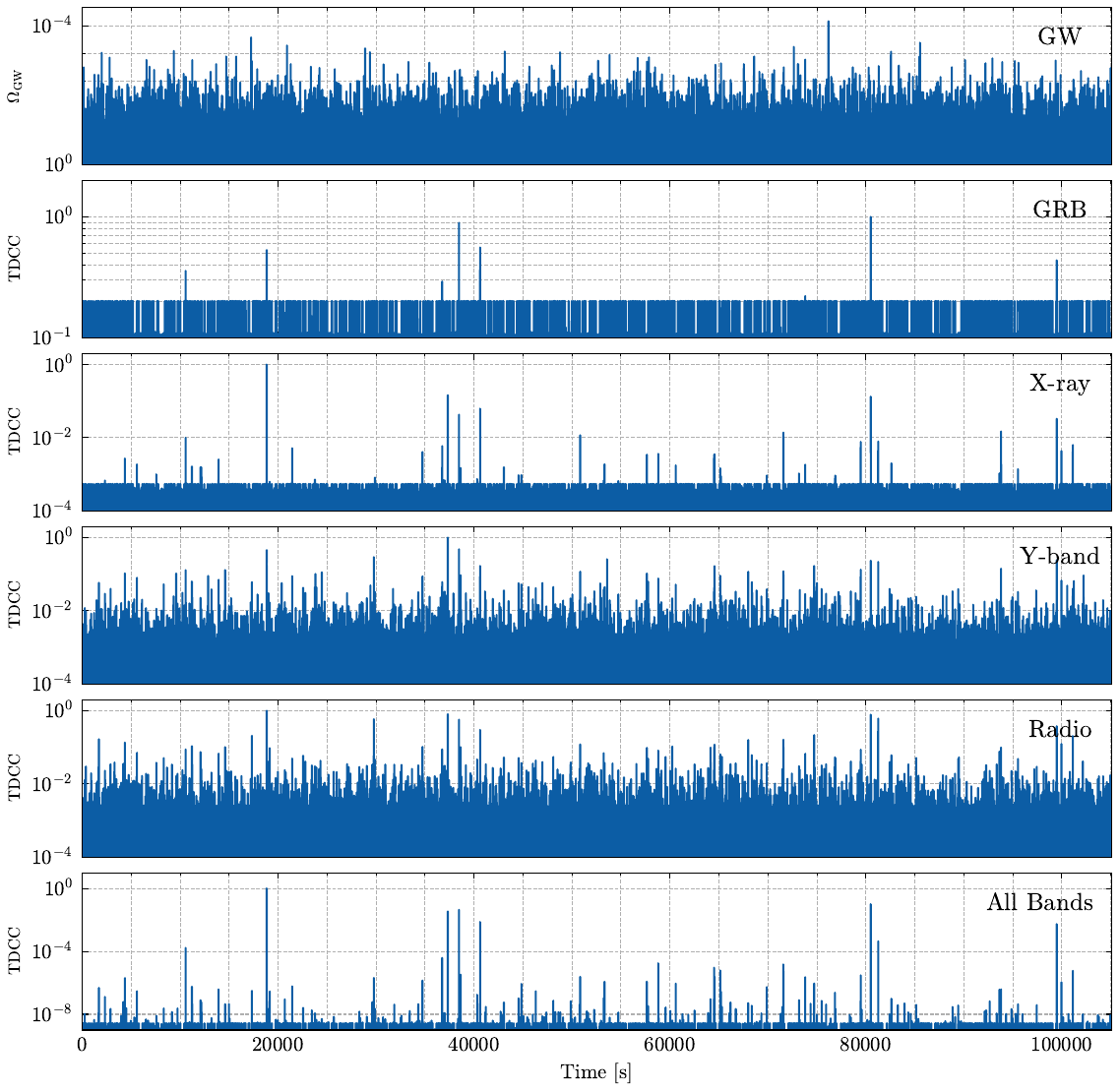}
    \caption{Top to bottom: $\Omega_{\text{GW}}(t)$; Normalized time-domain cross-correlation coefficient for correlation of  $\Omega_{\text{GW}}(t)$ with gamma-ray bursts, X-ray, Y-band, and radio waves. The bottom-most panel shows the combination of all bands. Negative correlations are not shown here for visual clarity.}
    \label{figp:200tdcc}
\end{figure*}

The results of the TDCC analysis with 
$\Delta t= 200$s are shown in Figure \ref{figp:200tdcc}. The figure spans a total observation window of 
$\sim10^5$s and presents the normalized TDCC coefficients between the SGWB and EM light curves, from gamma-ray to radio frequencies. Normalization ensures that the maximum value in each panel is unity, enabling a direct comparison across bands. For clarity, only positive correlations are displayed.

The top panel shows the TDCC for the frequency-integrated SGWB data. The subsequent panels present the TDCC results between the $\Omega_{\text{GW}}(t)$ and individual EM bands. In the GRB panel, sharp localized peaks appear over noise correlations. A similar observation is made for the case of X-rays. The lower frequency bands of optical (Y-band) and radio signals display some peaks that are temporally aligned with those observed in higher-energy bands. These two bands also encounter contamination from other uncorrelated sources and appear as false alarms (more on this in Section \ref{sec:FAR detect}). The final panel represents the combined TDCC. 
Peaks that were marginal in individual bands become more statistically significant when the multi-band information is combined. While some sharp peaks that were observed in one band diminish when combined with other bands. Therefore, combining multiple bands of the EM signal improves our detection capability, enhancing the detection of true events and rejecting false alarms.

The results demonstrate that the technique is capable of identifying statistically significant correlations between GW signals, submerged in detector noise, and EM observations in separate bands. Despite the presence of noise, the correlation remains discernible, validating the robustness of the approach. When the associations across all EM bands are collectively considered, the final panel highlights a consistent signal appearing across all messengers. This multi-messenger coincidence serves as compelling proof of the validity of this pipeline.

\subsection{TDCC with GW from unmodeled sources}
\begin{figure*}
    \centering
    \includegraphics[width=\textwidth]{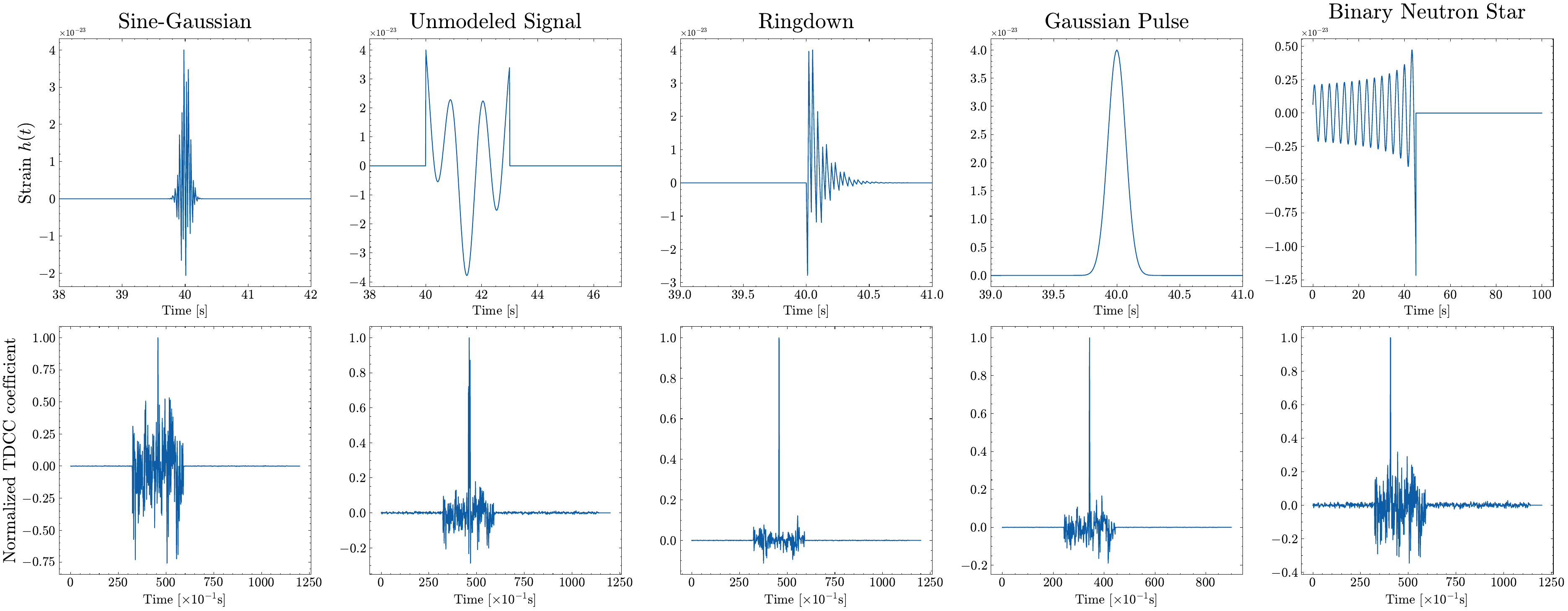}
    \caption{Top: Strains $h(t)$ vs. Time [s] for sine-Gaussian, unmodeled source, ringdown, Gaussian, and binary neutron star. Bottom: TDCC between GW flux (see Eq. \eqref{eqn:TDCC_ht}) of the corresponding waveforms in the top panel and a GRB signal with $\Delta t_{\nu}=4$s. The jitters in the TDCC coefficients arise from the detector noise.}
    \label{figp:gw_burst}
\end{figure*}

We apply the formalism to GW signals from unmodeled/unknown sources. We use the models described in Section \ref{sec:gw burst} to generate the GW signal and add detector noise with A+ sensitivity. We then perform TDCC of GW flux generated from these signals with a GRB signal. The models and the normalized TDCC with time domain strain data (see Eq. \eqref{eqn:TDCC_ht}) are shown in the top and bottom panels of Figure \ref{figp:gw_burst}, respectively. We have assumed a time delay of $\Delta t_{\nu}=4$s. The amplitude of the strain is chosen for this case as $10^{-24}$ m. The effectiveness of the correlation technique is influenced by both the amplitude of the GW signal and the assumed time delay between the GW and EM emissions. Higher amplitudes of signals yield better signal-to-noise ratios, enhancing the likelihood of detecting true correlations. Conversely, weak signals are more susceptible to being masked by detector noise and astrophysical background, increasing the chance of false positives. Additionally, the time delay between GW and EM signals, arising from intrinsic emission lags or propagation effects, also impacts the correlation. A wide or uncertain delay window can dilute the signal, but the TDCC method accommodates such variability by scanning across delay ranges. Despite these challenges, the formalism remains robust. Detector noise and background fluctuations do introduce variations in the correlation, yet the method consistently reveals clear associations between GW and EM signals. Crucially, the results confirm that the technique performs well for both modeled and unmodeled sources, establishing its model-agnostic nature and broad applicability in multi-messenger searches. The success of both strongly relies on the confident detection of EM signals in multiple bands, which helps in tracing back GW signals from the noise-dominated data.

\section{False Alarm Rate (FAR) Detection} \label{sec:FAR detect}
Many telescopes and observatories across the globe continuously monitor the transient EM sky over a wide range of frequencies, including gamma rays, X-rays, ultraviolet, optical, infrared, and radio bands. These facilities are designed to detect a variety of astrophysical transients, only some of which are expected to be associated with GW events. However, not all EM signals observed within the temporal or spatial proximity of a GW trigger originate from a common astrophysical source. In fact, some of these coincidental detections may arise from independent or unrelated processes in the universe. When such unassociated EM signals nevertheless exhibit a correlation with GW signals, particularly in a TDCC analysis, they can lead to a false alarm.

To illustrate, consider a scenario in which an X-ray signal, not physically associated with any GW event, happens to show a temporal correlation with a detected GW signal. If this X-ray event does not coincide with any signals in other EM bands, such as radio, optical, or GRBs, from the same region of the sky, it becomes increasingly likely that the observed correlation is spurious and is a false alarm. In contrast, a genuine multi-messenger astrophysical event is expected to produce coherent signals across multiple EM channels, all originating from the same sky location and temporally consistent with the GW event. The absence of such multi-wavelength confirmation undermines the credibility of the observed X-ray–GW correlation.

Therefore, estimating the FAR becomes a crucial component in evaluating the statistical significance of such observed coincidences. FAR quantifies the expected number of false correlations per year that could arise due to random temporal or spatial alignments between EM and GW signals. A low FAR strengthens the case for a true astrophysical association, whereas a high FAR signals the need for caution and further multi-wavelength verification. By rigorously evaluating the FAR, one can robustly differentiate between statistically coincidental alignments and physically meaningful multi-messenger events.

\begin{table}[h]
    \centering
    \begin{tabular}{ccccc}
    \hline
    \hline
       EM Band  & Detector & Event Rate [$\rm yr^{-1}$] & $z_{\text{max}}$ & SNR\\ \hline
       GRB   & Fermi-GBM & 120 & 1.0  & $>5\sigma$ \\ 
       Xray   & Swift-XRT & 125 & 0.2  & $>3\sigma$ \\  
       UVOIR (Y)  & Roman & 800 & 0.3 & $>3\sigma$ \\  
       Radio  & VLA/VLBI & 1000 & 1.0  & $>3\sigma$ \\ \hline
    \end{tabular}
    \caption{Description of EM signal injection, which are not associated with any GW merger event, according to bands up to a $z_{\text{max}}$ specific to the detectors, namely, Fermi-GBM \citep{fermi}, Swift-XRT \citep{SWIFT_GRB}, Roman \citep{roman}, VLA \citep{VLA}/VLBI \citep{vlbi}. The event rate is quantified per year.} 
    \label{tab:em_insertion}
\end{table}

To estimate the FAR from the distribution, we perform the TDCC between GW noise and EM signals obtained by injecting the above false counterpart events. We generate the GW background noise of A+ design sensitivity for a duration of one year with three different integration times of $100$s, $200$s, and $300$s.

The signal detected by the GW detectors depends on several variables, including the antenna detection functions ($F_+, F_\times$), inclination angle $i$, location of the source in the sky, and the orientation of the binary system with respect to the detector, among others. To quantify the strength of the GW signal relative to the background noise, we resort to using the following form of matched filtering SNR ($\rho$) \cite{Finn:1995ah},
\begin{equation}
    \rho = \frac{\Theta}{4} \left[ 4 \int_{f_{\text{min}}}^{f_{\text{max}}} h(f)^2 / S_n(f) ~df \right]^{1/2},
\end{equation}
where $\Theta^2 = 4 (F_+^2 (1+\cos^2 i) + 4 F_\times^2 \cos^2 i)$, with $\Theta$ following the distribution \cite{Finn:1995ah},
\begin{equation}
    P(\Theta) =
    \begin{cases}
    5\Theta(4-\Theta)^3/256, & 0 < \Theta < 4, \\
    0, & \text{otherwise}.
    \end{cases}
\end{equation}

For the purpose of our calculation, we consider the median value $\Theta \approx 1.25$. The signals in each band are observed by detectors which have a maximum redshift $z_{\text{max}}$ up to which they can observe. These detectors observe a variety of sources and events that result in emissions in a particular region. Quantifying event rates (per year), which are not necessarily related to a GW merger event, to an absolute number for EM emissions monitored by different detectors is largely uncertain; hence, we resort to some optimistic but physically motivated numbers for event rates for emissions in different bands \cite{Bissaldi:2019tpk, Ronchini:2022gwk}.  These EM signals, which are not associated with any GW merger event, are injected according to the event rates and detection capabilities of detectors as mentioned in Table \ref{tab:em_insertion}, considering flat cosmology with $H_0 = 67.7$ \si{km ~ s^{-1} Mpc^{-1}} and $\Omega_M = 0.31$ \citep{Planck_2020}. To make the case realistic, the event rates are associated with a particular detector in each band, where $z_{\text{max}}$ is the maximum redshift up to which the detector can observe. We also set an SNR cutoff to only include confident detections. These injected sources are distributed uniformly in the comoving volume, which refers to a redshift probability distribution of
\begin{equation}
    \frac{\mathrm{d} p(z)}{\mathrm{d} z} \propto \frac{1}{(1+z)} \frac{\mathrm{d} V_c}{\mathrm{d} z},
\end{equation}
where $V_c$ is the comoving volume. Each sampled event is then randomly assigned a time bin in a year-long data stream. This process ensures the construction a time-domain data stream of EM signals which are uncorrelated to each other and to any GW merger.

Specifically, we compute the cumulative distribution function (CDF) for the combined distribution of GW noise and EM signals. Our objective is to evaluate how likely it is to get a false alarm in the TDCC for an EM signal of a given strength (SNR). We analyze the distribution and the corresponding (1 - CDF) for the different GW events as given in Table \ref{tab:2}. We mention the source properties like $z_{\text{event}}$, $m_1$, $m_2$, and GW SNR, obtained via matched filtering technique, of inserted BNS events for which EM counterparts have been detected. The EM SNRs are scaled assuming a minimum SNR at $z_{\text{max}}$ of the given EM band as mentioned in Table \ref{tab:em_insertion}. The cells with null values show that the detector in that specific band cannot observe events at that redshift. 

For a given event, we compute the area under the curve of the distribution of the TDCC (between GW noise and EM signals) between the TDCC value of the event and infinity. This area is equivalent to the value of the complementary CDF at the point of TDCC of the event, which is the FAR for that event. We calculate the FAR by using a TDCC between GW and each band and also by using a combination of EM bands. Table \ref{tab:3} shows the results of an analysis done for integration time $\delta t = 300$s. The term \textit{min} corresponds to the minimum FAR value that an event can have for the given integration time. The \textit{min} is formally written as $\delta t / \text{(No. of seconds in a year)} $, and its value for $\delta t = 300$s is $9.5 \times 10^{-6}$. Here the null values again correspond to the redshift detection limit for a particular EM band (and in the combination). We repeat the same analysis for $\delta t = 200 \text{s} ~\text{and} ~100 \text{s}$ in Table \ref{tab:4} and Table \ref{tab:5}, respectively. The length of the total duration of the signal is kept at 1 year to reduce the computational cost involved in generating signals in different messengers. 

The results presented in this study demonstrate a clear and significant trend: the FAR decreases systematically as more EM bands are jointly analyzed in correlation with GW signals. This outcome holds across all tested integration times in the TDCC analysis. When multiple independent EM channels are required to show temporal and spatial coincidence with a GW event, the probability of a random correlation across all channels decreases sharply. This substantially reduces the likelihood of false alarms and strengthens the confidence in the detected multi-messenger event.

However, it is important to consider the observational limitations imposed by current EM observatories. Certain EM bands, particularly X-ray and UV, are constrained by relatively shallow detection depths due to instrumental sensitivity and absorption by interstellar and intergalactic media. As a result, they are often incapable of detecting transient signals originating from high-redshift sources. This inherently limits their utility in probing the distant universe, and their standalone use in GW–EM correlation studies may lead to higher FAR values, especially for sources at cosmological distances. In contrast, EM bands such as gamma rays and radio waves offer substantially better redshift coverage. GRBs, for instance, are among the most luminous events in the universe and can be detected up to high redshifts. Similarly, radio afterglows of compact object mergers can persist over long timescales and be observed over cosmological distances with modern radio arrays. When these high-redshift-capable bands are combined in a joint analysis, the detection probability of true counterparts increases, while the FAR is minimized due to the enhanced reliability of multi-band coincidence.
\begin{table*}[h]
    \centering
    \begin{tabular}{cccccccc}
    \hline
    \hline
       $z_{\text{event}}$  & $m_1 [M_{\odot}]$ & $m_2 [M_{\odot}]$ & GW SNR & GRB SNR & Xray SNR & UV SNR & Radio SNR\\ \hline
        0.07 & 1.11 & 1.00 & 2.59 & 2190.61 & 28.92 & 72.59 & 1314.36 \\
        0.12 & 1.83 & 1.16 & 1.93 & 696.73 & 9.2 & 23.09 & 418.04 \\ 
        0.23 & 1.98 & 1.16 & 1.04 & 166.03 & - & 5.50 & 99.62 \\ 
        0.45 & 1.78 & 1.42 & 0.59 & 34.99 & - & - & 20.99 \\ 
        0.67 & 2.32 & 1.68 & 0.49 & 13.39 & - & - & 8.03 \\
        0.98 & 2.01 & 1.47 & 0.31 & 5.26 & - & - & 3.15 \\ \hline
    \end{tabular}
    \caption{Source properties of inserted BNS multimessenger events with $z_{\text{event}}$, $m_1$ and $m_2$, and SNRs. GW SNR is the matched filtering SNR of the event, while the EM SNRs are scaled assuming minimum SNR at $z_{\text{max}}$ of the given EM band as mentioned in Table \ref{tab:em_insertion}. }
    \label{tab:2}
\end{table*}

\begin{table*}[h]
    \centering
    \begin{tabular}{cccccccccc}
    \hline
    \hline
    $z_{\text{event}}$  & G$\gamma$ &  GX & GU & GR& G$\gamma$X & G$\gamma$XU & G$\gamma$XUR & G$\gamma$U & G$\gamma$R\\ \hline
    0.07 & $\leq$ min & $2 \times 10^{-5}$ & $\leq$ min & $\leq$ min & $\leq$ min & $\leq$ min & $\leq$ min & $\leq$ min & $\leq$ min\\
    0.12 & $\leq$ min & $1.4 \times 10^{-4}$ & $2.9 \times 10^{-4}$ & $\leq$ min & $\leq$ min & $\leq$ min & $\leq$ min & $\leq$ min & $\leq$ min \\ 
    0.23 & $2 \times 10^{-5}$ & - & $4.1 \times 10^{-3}$ & $10^{-4}$ & - & - & - & $\leq$ min & $\leq$ min \\
    0.45 & $1.2 \times 10^{-4}$ & - & - & $6.6 \times 10^{-4}$ & - & - & - & - & $\leq$ min \\
    0.67 & $2.8 \times 10^{-4}$ & - & - & $1.4 \times 10^{-3}$ & - & - & - & - & $\leq$ min \\
    0.98 & $1.6 \times 10^{-3}$ & - & - & $1.4 \times 10^{-2}$ & - & - & - & - & $\leq$ min \\ \hline
    \end{tabular}
    \caption{FAR (per year) for different combinations of GW and EM signals for $\Delta t_{\text{int}} = 300s$. The combination codes are as follows: G is GW, $\gamma$ is gamma-ray, X denotes X-ray, U is the collection of UV/optical/IR bands, while R signifies radio. For this $\Delta t_{\text{int}}$, min $ \leq 9.5 \times 10^{-6}$.}
    \label{tab:3}
\end{table*}

\begin{table*}[h]
    \centering
    \begin{tabular}{cccccccccc}
    \hline
    \hline
    $z_{\text{event}}$  & G$\gamma$ &  GX & GU & GR& G$\gamma$X & G$\gamma$XU & G$\gamma$XUR & G$\gamma$U & G$\gamma$R\\ \hline
    0.07 & $\leq$ min & $\leq$ min  & $2 \times 10^{-5}$ & $\leq$ min & $\leq$ min & $\leq$ min & $\leq$ min & $\leq$ min & $\leq$ min \\ 
    0.12 & $\leq$ min & $1.7 \times 10^{-4}$ & $1.6 \times 10^{-4}$ & $2 \times 10^{-5}$ & $\leq$ min & $\leq$ min & $\leq$ min & $\leq$ min & $\leq$ min \\ 
    0.23 & $\leq$ min & - & $1.7 \times 10^{-4}$ & $9 \times 10^{-5}$ & - & - & - & $\leq$ min & $\leq$ min \\ 
    0.45 & $8 \times 10^{-5}$ & - & - & $5.3 \times 10^{-4}$ & - & - & - & - & $\leq$ min \\ 
    0.67 & $1.6 \times 10^{-4}$ & - & - & $1.3 \times 10^{-3}$ & - & - & - & - & $\leq$ min \\ 
    0.98 & $3.5 \times 10^{-3}$ & - & - & $2.5 \times 10^{-2}$ & - & - & - & - & $\leq$ min \\ \hline
    \end{tabular}
    \caption{FAR (per year) for different combinations of GW and EM signals for $\Delta t_{\text{int}} = 200s$. The combination codes are as follows: G is GW, $\gamma$ is gamma-ray, X denotes X-ray, U is the collection of UV/optical/IR bands, and R signifies radio. For this $\Delta t_{\text{int}}$, min $ \leq 6.3 \times 10^{-6}$.}
    \label{tab:4}
\end{table*}

\begin{table*}[h]
    \centering
    \begin{tabular}{cccccccccc}
    \hline
    \hline
       $z_{\text{event}}$  & G$\gamma$ &  GX & GU & GR& G$\gamma$X & G$\gamma$XU & G$\gamma$XUR & G$\gamma$U & G$\gamma$R\\ \hline
        0.07 & $\leq$ min & $10^{-5}$ & $7 \times 10^{-5}$ & $10^{-5}$ & $\leq$ min  & $\leq$ min  & $\leq$ min & $\leq$ min & $\leq$ min \\ 
0.12  & $\leq$ min & $1.2 \times 10^{-3}$ & $1.9 \times 10^{-3}$ & $2 \times 10^{-5}$ & $\leq$ min & $\leq$ min & $\leq$ min & $\leq$ min & $\leq$ min \\ 
0.23 & $10^{-5}$ & - & $1.1 \times 10^{-1}$ & $9 \times 10^{-5}$ & - & - & - & $\leq$ min & $\leq$ min \\ 
0.45 & $5 \times 10^{-5}$ & - & - & $3.5 \times 10^{-4}$ & - & - & - & - & $\leq$ min \\ 
0.67 & $2.2 \times 10^{-4}$ & - & - & $1.8 \times 10^{-3}$ & - & - & - & - & $\leq$ min \\ 
0.98 & $1.2 \times 10^{-2}$ & - & - & $4.5 \times 10^{-2}$ & - & - & - & - & $\leq$ min \\ \hline
    \end{tabular}
    \caption{FAR (per year) for different combinations of GW and EM signals for $\Delta t_{\text{int}} = 100s$. The combination codes are as follows: G is GW, $\gamma$ is gamma-ray, X denotes X-ray, U is the collection of UV/optical/IR bands, and R signifies radio. For this $\Delta t_{\text{int}}$, min $ \leq 3.2 \times 10^{-6}$.}
    \label{tab:5}
\end{table*}

In a similar fashion, the FAR analysis can be done for the case of signals from unmodeled sources. However, since the event rates for signals from unmodeled sources are unknown, a concrete number cannot be estimated. GW signature can also be positively anticipated from such events in the future. FAR calculations will demonstrate its minimum value for such events.

{\section{Estimated Event Detection Rate Using TDCC}
\label{sec:estimated-event-rate}

We outline a method for estimating the number of sub-threshold BNS events recoverable via TDCC under realistic detector sensitivity and observational constraints. Using the simulated SGWB signal (refer to section \ref{sec:popmodels} for details of the considered population model) with signal integration timescale $\Delta t =200 \rm s$ assuming the A+ sensitivity of the detector \citep{barsotti2018a+}. We focus exclusively on the GRB counterpart as the primary EM lookout for the following reasons. First, GRBs are the prompt, high-energy signature of BNS mergers. Second, Fermi-GBM monitors a large fraction of the sky continuously, making it uniquely suited for untriggered searches. Third, and most importantly, the only confirmed multimessenger BNS event to date (GW170817) was detected precisely through this prompt GRB association \cite{abbott_gw170817:_2017}. Follow-up observations in X-ray, UV, and radio require prior sky localisation from either the GW network or the GRB trigger, and their smaller fields of view make blind detection from an isotropic population impractical. For GRB, motivated by the only observation of GW170817, the jet half-opening angle is fixed to $\theta_j = 5^{\circ}$, consistent with the upper limit from \cite{kasliwal_illuminating_2017}. We consider two scenarios: an on-axis jet, in which a GRB is detectable only if the observer lies within the jet core, and an off-axis jet, as GW170817, in which structured jet emission remains detectable out to a viewing angle of $\Theta \leq 28^\circ$ \cite{LIGOScientific:2017zic}.

For each simulated merger event in the background signal, we compute time-domain energy light curves for GRB (as described in Appendix \ref{subsec:grb170817a}), and the received energy flux at the detector is computed from the luminosity distance of each event. The GRB SNR is then computed by integrating the flux over the signal duration $\Delta t = 200\,\rm s$ and dividing by the instrument noise level appropriate for \textit{Fermi-GBM} \cite{fermi, LIGOScientific:2017zic}.  We perform TDCC against the simulated GW background to compute the TDCC SNR for each event. 

\subsection{Detection Criteria}
Detection is determined by a set of four binary condition vectors denoted by $\{\mathbf{d}_i\}$, each encoding a distinct physical or instrumental requirement that must be satisfied for an event to be counted as detected. Each vector has entries $d_i^{(k)} \in \{0, 1\}$ for the $k$-th event, where $d_i^{(k)} = 1$ indicates that condition $i$ is satisfied and $d_i^{(k)} = 0$ indicates it is not. The total number of detected events is then determined by the product $N_\mathrm{det} = \sum_{k=1}^N \prod_i \mathbf{d}_i^{(k)}$, where the product ensures that all conditions must be simultaneously satisfied. The four conditions are evaluated in strict sequential order; if any condition returns 0, the pipeline stops for that event, and it contributes zero to $N_\mathrm{det}$.

The four detection conditions are applied in the following order. First, $\mathbf{d}_1$ encodes the jet orientation and is evaluated first since the GRB signal amplitude depends directly on whether the jet is pointed toward the observer. The inclination angle $i$ is drawn isotropically for all merger events, i.e. $\cos i \sim \mathcal{U}(-1, 1)$. For the on-axis case, a GRB is detectable only if $\cos i \geq \cos\theta_j$, which reflects the solid angle of a two-sided jet cone. For the off-axis case, the condition is relaxed to $\Theta \leq 28^\circ$, assuming a value similar to GW170817. If $\mathbf{d}_1=0$, no detection is possible and the pipeline stops for that event.

Second, $\mathbf{d}_2$ encodes instrument visibility by accounting for the finite sky coverage and operational duty cycle of \textit{Fermi-GBM} \cite{fermi}. The joint probability that a given event is both within the field of view (FOV) of the instrument and observed during an active window is written as $p = f_{\rm sky} \times \eta$, where $f_{\rm sky} = \Omega_{\rm FOV} / 4\pi$ is the fraction of sky covered by the instrument, $\Omega_{\rm FOV} = 8 \rm ~sr$ \cite{galaxies6040117} is the opening-angle equivalent of the instruments FOV, and $\eta = 0.6$ \footnote{\url{https://fermi.gsfc.nasa.gov/ssc/observations/types/grbs/}} is the duty cycle (the fraction of time the instrument is collecting science data, reduced from unity primarily by passages through the South Atlantic Anomaly and Earth occultation). A detection flag, $\mathbf{d}_2^{(k)} \in \{ 0,1\}$, is drawn from a Bernoulli distribution with this probability. 

It is important to clarify that the jet opening angle $\theta_j$ entering $\mathbf{d}_1$ and the instrument FOV $\Omega_{\rm FOV}$ entering $\mathbf{d}_2$ are physically distinct quantities. The jet opening angle $\theta_j$ is a property of the \textit{source}: it describes the angular extent of the outflow from the merger and determines whether the emitted GRB radiation is beamed toward the observer at all. The field of view $\Omega_{\rm FOV}$, by contrast, is a property of the \textit{instrument}: it describes the fraction of the sky that \textit{Fermi-GBM} is sensitive to at any given time, independent of the source geometry.

Third, $\mathbf{d}_3$ encodes the SNR threshold for GRB: since the signal amplitude is conditioned on $\mathbf{d}_1$, the SNR entering $\mathbf{d}_1$ is zero for misaligned jets, and $\mathbf{d}_1$ is evaluated only for events passing both $\mathbf{d}_1$ and $\mathbf{d}_2$. The GRB SNR is computed from the modeled prompt emission light curve as described in Appendix \ref{subsec:grb170817a}, scaled by the luminosity distance. We condition a detection to $\mathrm{SNR}_\mathrm{GRB} > 5\sigma$. 

Finally, $\mathbf{d}_4$ flags events for which the TDCC SNR also exceeds the same $\mathrm{SNR_{TDCC}} > 5\sigma$ threshold. 

\subsection{Results}
{The results for the expected detection rates for BNS events over one year of simulated data (with a local merger rate $R(0) = 100$ \si{Gpc^{-3}yr^{-1}} for BNS, with a total of 9802 BNS mergers in year for $z \leq 1.0$) are summarized here for the on-axis and off-axis cases, assuming GW170817 BNS merger and jet property. In the on-axis scenario, the strict jet alignment condition ($\mathbf{d}_1$) $\cos i \geq \cos(5^\circ)$ yields 18 BNS mergers. Of these, 8 BNS pass the instrument visibility cut $\mathbf{d}_2$ and 6 BNS events have $\rm SNR_{\rm GRB}$ exceeding $5\sigma$ ($\mathbf{d}_3$). After the TDCC significance cut ($\mathbf{d}_4$), we are left with 4 BNS mergers, of which the farthest event lies at a redshift $z=0.62$ from our mock simulation. 

In the off-axis scenario, like GW170817, the condition $\mathbf{d}_1$ yields 572 BNS events with lines of sight within the detectable cone. Of these, 217 BNS survive the \textit{Fermi-GBM} visibility cut $\mathbf{d}_2$, and 4 BNS mergers have EM SNR exceeding the $5 \sigma$ threshold ($\mathbf{d}_3$). After applying the TDCC significance cut ($\mathbf{d}_4$), we find a total of $N_\mathrm{det} = 3$ detected GW170817-like BNS events per year with A+ sensitivity. The farthest off-axis detection at $z=0.018$, which is consistent with the maximum GBM detection distance for GRB170817-like events \cite{LIGOScientific:2017zic}. }

Several caveats apply to this estimate. First, NSBH mergers are not included in this analysis because jet formation in NSBH systems is quite uncertain. The jet formation in NSBH systems requires tidal disruption of the neutron star by the BH, which occurs only for low mass ratios and high prograde BH spins (see \cite{Clarke:2024idd} and references therein). For the bulk of the NSBH population, the neutron star may be swallowed whole without forming an accretion disk, producing no prompt GRB emission \cite{PhysRevD.98.081501, Biscoveanu:2023eyp}. Including NSBH events without explicitly modeling the tidal disruption condition is highly assumption-driven and would therefore overestimate the GRB detection rate. Secondly, both the jet opening angle $\theta_j$ and the maximum detectable viewing angle $\Theta$ are informed solely by GW170817, the only confirmed BNS multimessenger event to date. The jet half-opening angle $\theta_j < 5^\circ$ is constrained from jet modeling \citep{kasliwal_illuminating_2017}, while the viewing angle upper limit $\Theta \leq 28^\circ$ is derived from a combination of GW parameter estimation and the Hubble flow velocity of the host galaxy NGC4993 \cite{LIGOScientific:2017zic}. These two quantities are physically and observationally distinct, and adopting them as fixed parameters for a cosmological population of BNS mergers neglects any astrophysical variations and hence can introduce additional uncertainty. Moreover, the merger rate considered in the analysis is also uncertain. Though it agrees with the current rates from the current LVK analysis \cite{LIGOScientific:2025pvj}, there is a large measurement uncertainty. An increase (or decrease) in the merger rate of BNS would lead to a corresponding increase in the number of detectable events from SGWB using TDCC, and hence is another potential source of uncertainty in the above-mentioned detection rates based on the mock simulation.

A successful detection of the BNS merger event using the GRB detection from a sub-threshold BNS merger via TDCC would immediately enable targeted follow-up across the EM spectrum with the GRB sky localisation from \textit{Fermi-GBM}
enabling X-ray follow-up with \textit{Swift-XRT}, UV/optical observations with the Nancy Grace Roman Space Telescope, and radio afterglow monitoring with VLA/VLBI, for example. Each subsequent band probes a different physical component of the merger. The joint detection of GRB, kilonova, and afterglow emission from a sub-threshold event would therefore not only confirm the merger origin of the signal but would also provide independent constraints on various other properties of the system.
 }

\section{Conclusion}\label{sec:Conclusion}

In this paper, we introduced the $\mathbf{MC^2}$, \textbf{M}ulti-messenger \textbf{C}ross-\textbf{C}orrelation analysis pipeline to detect GW events using the sporadic GW background and EM signal. The current detection capabilities of LVK restrict us to a limited observation horizon for GW. We build a formalism that performs the time-domain cross-correlation between the GW background signal and the EM signal to look for potential counterparts of the GW events that are shrouded in the background. The pipeline is developed to utilize multi-messenger coincidences, allowing us to extract information even from GW signals that are individually undetectable with matched filtering. We demonstrated the effectiveness of $\mathbf{MC^2}$ using simulated data across multiple EM bands, showing a significant reduction in false alarm rates when multi-band observations are combined. The framework successfully identifies correlations between GW and EM signals, even in the presence of noise. The framework leverages the high detection horizons of EM signals across multiple bands, including gamma rays, X-rays, UV/optical/IR, and radio. The multi-band approach not only facilitates the detection of very faint GW events but also makes possible the precise localization and redshift determination from their EM counterparts. {To quantify the practical utility of TDCC under realistic observational conditions, we estimate the expected detection rate of sub-threshold BNS events by folding in the sky coverage, duty cycle, and detection thresholds of \textit{Fermi-GBM}. This translates the theoretical sensitivity of TDCC into a concrete, instrument-aware prediction for the number of multimessenger events recoverable per year.} The results demonstrate that the pipeline is effective even for unmodeled sources, making it a versatile tool for detecting and characterizing a wide range of astrophysical phenomena.

The applicability of the proposed cross-correlation technique does not depend on continuous or dedicated monitoring by any single electromagnetic survey. Instead, the method can be applied using intermittent, survey-specific electromagnetic observations, provided there is sufficient temporal overlap with the expected EM emission time window after the GW merger. Here, the correlation between GRB signals will be crucial, as they will be used to send triggers to other EM missions.

At present, the $\mathbf{MC^2}$ framework provides a promising approach for detecting GW events through multi-messenger cross-correlation techniques. For a study of this nature, data from the existing EM missions like the \textit{Fermi Gamma-ray Space Telescope} \cite{fermi}, which monitors in the gamma-bands; \textit{Swift} \cite{swift}, SXT aboard the \textit{AstroSat} \cite{astrosat}, observe the X-rays; \textit{Zwicky Transient Facility} (ZTF) \cite{ztf}, the \textit{Dark Energy Camera} (DECam) \cite{decam}, UVIT/OPT aboard the \textit{AstroSat} \cite{astrosat} look at the infrared and optical sky and radio observations from facilities such as the \textit{upgraded Giant Metrewave Radio Telescope} (uGMRT) \cite{ugmrt}, the \textit{Very Large Array} (VLA) \cite{VLA}, will be very useful.

Along with this, advent of data from upcoming EM observatories across multiple wavelengths, such as, \textit{Athena} \cite{Athena_2020,bavdaz2017athena} in X-ray, \textit{Vera C. Rubin Observatory} \cite{hambleton2022rubinobservatorylssttransients,lsst}, the \textit{Nancy Grace Roman Space Telescope} \cite{hounsell2023romanccswhitepaper,roman}, the\textit{ Gravitational-wave Optical Transient Observer} (GOTO) \cite{goto}, \textit{ULTRASAT} \cite{ultrasat},  and \textit{Euclid} \citep{laureijs2010euclid} in infrared and optical band and the \textit{Square Kilometre Array} (SKA) \citep{mcpherson2018square} in radio band, will further enhance the effectiveness of this approach.

Moreover, the framework is naturally suited for application to archival datasets, where GW data can be cross-correlated with EM observations accumulated over extended periods. This is especially relevant for transient phenomena such as GRB, which are short-lived, while their associated X-ray, optical, and radio afterglows can emerge on much longer timescales. Available archival EM data, therefore, provide a unique opportunity to retrospectively search for correlated signatures that may not have been identifiable in real time, allowing weak or sub-threshold GW signals associated with past events to be statistically recovered. These advancements will enable the $\mathbf{MC^2}$ formalism to detect faint and sub-threshold GW signals, opening new possibilities for studying compact object mergers.

\begin{acknowledgments}
{We thank the referee for their constructive and insightful comments, in particular for suggesting the addition of expected detection rates, which have helped improve the quality and clarity of this paper.} The authors are grateful to Aniruddha Chakraborty for reviewing the manuscript as a part of the LSC Presentation and Publication procedure and providing useful comments. This work is a part of the \texttt{⟨data|theory⟩ Universe-Lab}, supported by the TIFR and the Department of Atomic Energy, Government of India (GoI). HJK is funded by the INSPIRE Scholarship for Higher Education, Department of Science and Technology (DST), GoI. HJK was also partially funded during the duration of this work by the National Initiative for Undergraduate Sciences (NIUS) program conducted by the Homi Bhabha Centre for Science Education (HBCSE), Mumbai, and acknowledges the support of the Department of Atomic Energy, GoI, under Project Identification No. RTI4001. The authors express gratitude to the \texttt{⟨data|theory⟩ Universe-Lab}'s cluster for meeting the computational needs. We thank the LIGO-Virgo-KAGRA Scientific Collaboration for providing noise curves. LIGO, funded by the U.S. National Science Foundation (NSF), and Virgo, supported by the French CNRS, Italian INFN, and Dutch Nikhef, along with contributions from Polish and Hungarian institutes. The research leverages data and software from the Gravitational Wave Open Science Center, a service provided by LIGO Laboratory, the LIGO Scientific Collaboration, Virgo Collaboration, and KAGRA. Advanced LIGO's construction and operation receive support from STFC of the UK, Max-Planck Society (MPS), and the State of Niedersachsen/Germany, with additional backing from the Australian Research Council. Virgo, affiliated with the European Gravitational Observatory (EGO), secures funding through contributions from various European institutions. Meanwhile, KAGRA's construction and operation are funded by MEXT, JSPS, NRF, MSIT, AS, and MoST. This material is based upon work supported by NSF’s LIGO Laboratory, which is a major facility fully funded by the National Science Foundation. This material is based upon work supported by the LIGO Laboratory, which is a major facility fully funded by the National Science Foundation. The authors acknowledge the use of the following packages for this work: \texttt{Astropy} \cite{Astropy}, \texttt{NumPy} \cite{NumPy}, \texttt{Matplotlib} \cite{Matplotlib}, \texttt{Pandas} \cite{Pandas,Pandas2}, \texttt{SciPy} \cite{SciPy}, \texttt{healpy} \cite{healpy}, and \texttt{HEALPix} \cite{HEALPix}.
\end{acknowledgments}

\appendix

\section{Parametric modeling of EM counterparts from binary neutron stars}\label{sec:EMmodel}

\subsection{Gamma Ray Bursts} \label{subsec:grb170817a}
 
The Fermi Gamma-ray Burst Monitor (GBM) \cite{Goldstein_2017} detected a short gamma-ray burst, henceforth dubbed GRB170817A, 1.7 s after GW170817. GRB170817A is two orders of magnitude closer and two to six orders less energetic than other SGRBs detected so far. The basic properties like the peak energy, spectral slope, and duration of the main peak are in line with existing prompt emission models (e.g., dissipative photosphere, \cite{Rees_2005}; internal shocks, \cite{Rees_1994}). In addition to the main peak, which lasted for around 0.5 seconds, there was a softer secondary emission that lasted for approximately 1.1 seconds. 

Employing these observations of emission and the event properties, we adapt the pulse profile described in \cite{Norris_1996}, where the pulse shape is given by
\begin{equation} \label{eqn:grb_model}
    I_{\text{GRB}}(t) =
    \begin{cases}
        A e^{- \left( \frac{t_{\text{peak}} - t}{\sigma_{\text{rise}}} \right)^\nu } & \text{, for $t < t_{\text{peak}}$,}\\
        A e^{- \left( \frac{t - t_{\text{peak}}}{\sigma_{\text{decay}}} \right)^\nu } & \text{, for $t > t_{\text{peak}}$.}
    \end{cases}
\end{equation}
here $A$ is the amplitude at the peak time of the pulse, $t_{\text{peak}}$, $\sigma_{\text{rise}}$ and $\sigma_{\text{decay}}$ are the characteristic rise and decay times of the pulse respectively.
\begin{figure*}
\centering
\includegraphics[width=.46\linewidth]{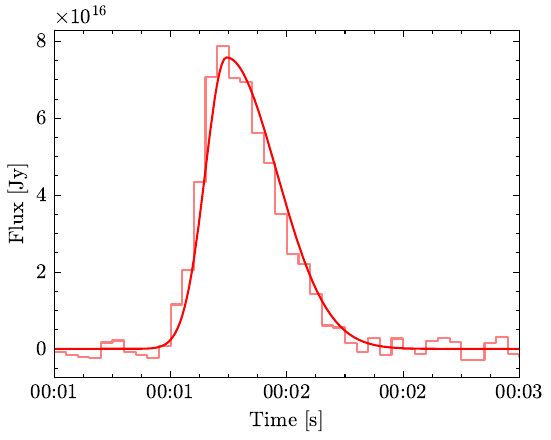}
\qquad
\includegraphics[width=.46\linewidth]{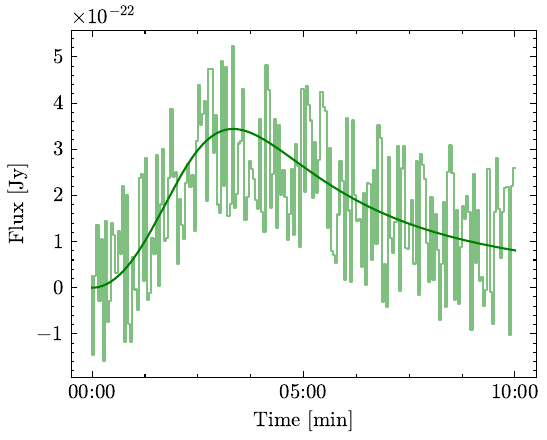}
\caption{Simulated signal of a gamma ray (left) and X-ray (right) depicted in a flux [Jy] vs. Time [s] plot. The solid line highlights the waveform model, while the step plot is the noise-added signal.}
\label{figp:gamma-xray}
\end{figure*}

\subsection{X-rays} \label{subsec:xrays}

Nine days post-trigger of GW170817, \cite{Troja_2017} reported the discovery of the X-ray counterpart with Chandra X-ray Observatory. A 50 ks exposure observation detected significant X-ray emission at the same position as the optical/IR counterpart \cite{Troja_2017}. The observed X-ray flux implies an isotropic luminosity of $9 \times 10^{38}$ \si{erg.s^{-1}} if located in NGC 4993 at a distance of $~40$ Mpc. Fifteen days post-trigger Chandra observations performed between 1st and 2nd Sept 2017, two additional 50 ks observations were made, which confirmed the continued presence of X-ray emission with a slight increase in luminosity to $L_{\text{X,iso}} ~ 1.1 \times 10^{39}$ \si{erg.s^{-1}}. Based on the first of these two observations, \cite{Fong_2017_GCN} reported the detection of the X-ray counterpart and the presence of an additional X-ray point source in the near vicinity. \cite{Margutti_2017} and \cite{Troja_2017_GCNe} reported a flux of $4.5 \times 10^{-15} $\si{erg.cm^{-2}.s^{-1}} for the X-ray counterpart. One day later, \cite{Haggard_2017_GCNb} reported another deep observation showing continued distinct X-ray emission coincident with SSS17a/AT2017gfo, NGC 4993, and the additional point source \citep{Haggard_2017,Haggard_2017_GCNb}. We adopt a generic profile modeling the X-ray signal as detected by \cite{Beuermann_1999} and \cite{Troja_2019}, 

\begin{equation} \label{eqn:xray_model}
    I_{\text{xray}}(t) = A \left( \frac{\nu_{\text{xray}}}{\nu_0} \right)^{-b} \left( \left( \frac{t}{t_{\text{peak}}} \right)^{-a_1 \kappa} + \left(\frac{t}{t_{\text{peak}}}\right)^{a_2 \kappa} \right)^{-1/\kappa},
\end{equation}
where $A$ is the amplitude of the signal, $\nu_{\text{xray}}$ is the frequency of the X-ray signal, between $3 \times 10^{16}$ \si{Hz} and $3 \times 10^{19}$ \si{Hz}, and $\nu_0 = 10^{7}$ \si{Hz} is the pivot point. The first term encapsulates the frequency dependence of the X-ray emission using a power law with index $\beta = 0.585$. The peak time of the emission is denoted by $t_{\text{peak}}$, $a_1=0.9$ and $a_2=2.0$ are the decay and rise parameters, respectively, with $\kappa=2$ the smoothness parameter. The right panel of Figure \ref{figp:gamma-xray} depicts the simulated waveform model. X-ray observations of GRB afterglows are important to constrain the geometry of the outflow, its energy output, and the orientation of the system with respect to the observers’ line of sight.

\subsection{UV/Optical/IR} \label{subsec:uvoir}
At 11.4 hr post-merger, detection of a bright optical transient was reported close to the nucleus of NGC4993 at redshift $z=0.0098$ \citep{Soares-Santos_2017}, consistent with the distance of $40 \pm 8$ Mpc of GW170817. The detected transients had magnitudes $i = 17.30$ mag and $z = 17.43$ mag. These luminosity magnitudes tell us that the source might be a kilonova \citep{Metzger_2017}.

The parameters obtained from the observed light curves of the optical counterparts of GW170817 in six filters are shown in Table \ref{tab:uvoir}. The initial decay of the source was observed as $m \propto t^{-\alpha}$ \citep{Soares-Santos_2017}, where $\alpha$ is the decay parameter and $R_V$ characterizes the interstellar extinction from the NIR to the far-UV spectral region. The decay in the first few days is consistent with a peak luminosity near $t=1$ day. 
\begin{table}[h!] 
    \centering
    \begin{tabular}{c c c c c c c}
    \hline
    \hline
    Band  &  u & g & r & i & z & Y\\
    \hline
    Decay ($\alpha$) & 2.0 & 1.4 & 0.7 & 0.5 & 0.5 & 0.4\\ 
    $R_V$ & 3.963 &  3.186 &  2.140&  1.569 & 1.196& 1.048\\
    \hline
    \end{tabular}
    \caption{Decay of light in different bands with $\alpha$ is the decay parameter, and $R_V$ characterizes the interstellar extinction from the NIR to the far-UV spectral region.}
    \label{tab:uvoir} 
\end{table}

\begin{figure*}
\centering
\includegraphics[width=.46\linewidth]{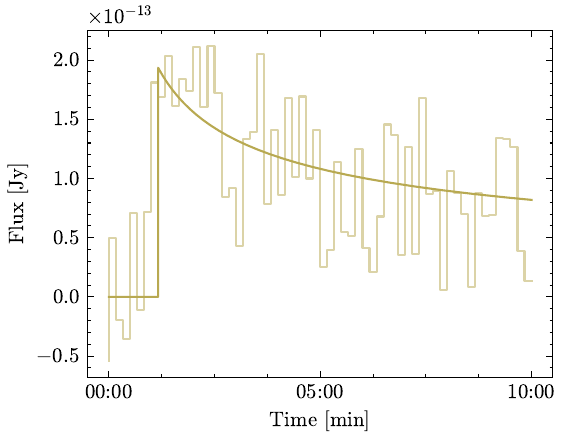}
\qquad
\includegraphics[width=.46\linewidth]{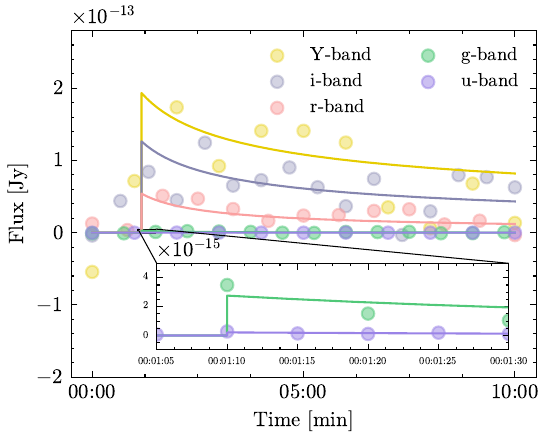}
\caption{Simulated fluxes of a Y-band (left) and of u, g, r, i, and Y-band signals depicted in a Flux [Jy] vs. Time [s] plot. The solid line highlights the flux model, while the step plot is the noise-added signal. The scatter plot in the right panel is the noise-added signal.}
\label{figp:y-uvoir}
\end{figure*}

We model our optical counterparts as follows
\begin{equation} \label{eqn:uvoir_model}
    m(t) = 
    \begin{cases}
        0 & t < t^{\text{UVOIR}}_0\\
        A t^{- \alpha} & t \geq t^{\text{UVOIR}}_0,
    \end{cases}
\end{equation}
where $t^{\text{UVOIR}}_0$ is the time of trigger of the UV/optical/IR counterpart. The simulated waveform model for the Y-band is shown in the left panel of Figure \ref{figp:y-uvoir}, while the right panel compares the strength and decay rates of the u, i, r, g, z, and Y band.

\subsection{Radio Waves} \label{subsec:radio}

The first radio observations of the optical transient SSS17a/AT2017gfo’s location (NGC4993) were started on August 18, 13.5 hr post-merger with the Karl G. Jansky Very Large Array (VLA) \citep{Alexander_2017_GCN}. The first radio peak detection was made with the VLA on 2017 September 2 and 3 at two different frequencies ($\sim 3$ GHz and $\sim 6$ GHz) via two independent observations \citep{Mooley_2017_GCN,Corsi_2017_GCN}. An evolving transient was confirmed by \citep{Hallinan_2017,Corsi_2017_GCN,Mooley_2017_GCN} through repeated detections scanning multiple frequencies. 

\begin{figure}
\centering
\includegraphics[width=\linewidth]{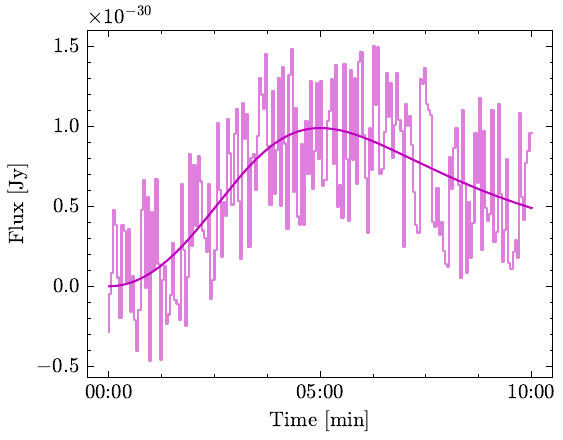}
\caption{Simulated flux of a radio signal depicted in a Flux [Jy] vs. Time [s] plot. The solid line highlights the waveform model, while the step plot is the noise-added signal.}
\label{figp:radio-ccsn}
\end{figure}

Radio and X-ray follow a strongly similar model of emission. We adopt the same generic profile as was used for X-rays in \eqref{eqn:xray_model} 
\begin{equation} \label{eqn:radio_model}
    I_{\text{radio}}(t) = A \left( \frac{\nu_{\text{radio}}}{\nu_0} \right)^{-b} \left( \left( \frac{t}{t_{\text{peak}}} \right)^{-a_1 \kappa} + \left(\frac{t}{t_{\text{peak}}}\right)^{a_2 \kappa} \right)^{-1/\kappa},
\end{equation}
where $A$ is the amplitude of the signal, $\nu_{\text{radio}}$ is the frequency of the radio signal, between $3 \times 10^{11}$ \si{Hz} and $ 9 \times 10^{3}$ \si{Hz}, and $\nu_0 = 10^{7}$ \si{Hz} is the pivot point. The first term encapsulates the frequency dependence of the radio emission using a power law with index $\beta = 0.585$. The peak time of the emission is denoted by $t_{\text{peak}}$, $a_1=0.9$ and $a_2=2.0$ are the decay and rise parameters, respectively, with $\kappa=2$ the smoothness parameter. Figure \ref{figp:radio-ccsn} shows the simulated waveform model.

While the proposed correlation technique is designed to be model-agnostic, it is important to acknowledge certain caveats associated with modeling EM signals. The temporal and spectral properties of EM emissions, such as their duration, delay, intensity profiles, and emission mechanisms, can vary significantly depending on the nature of the astrophysical source (e.g., binary neutron star merger or black hole–neutron star merger). Many of these emissions, particularly in the X-ray and radio bands, are subject to considerable modeling uncertainties, including unknown delay times relative to the GW signal, propagation effects, and poorly constrained source physics \cite{Piro:2018bpl,Nakar:2011cw}.

These uncertainties can introduce ambiguities in the expected EM waveform shapes or their temporal alignment with the GW signal, which may affect the performance of the TDCC technique. For instance, if the true EM signal is significantly delayed or spread in time compared to modeled expectations, the correlation may weaken or fall outside the tested time-delay window, resulting in missed associations or underestimated significance. Moreover, the use of simplified or idealized EM signals is often necessary due to limited observational data but may not capture the diversity of real astrophysical scenarios. This could bias the interpretation of correlation strength or produce spurious matches in the presence of noise.

Nevertheless, the technique remains robust as it does not rely on strict waveform matching but rather on the statistical alignment of energy flux over time. However, improved EM modeling and better constraints on time-delay distributions will further enhance the sensitivity and reliability of multi-messenger searches using this approach.

\section{Implementation and Effectiveness} \label{sec:implementation}
The time delay, $\Delta t_{\nu}$, and $\delta t$, the small interval over which we are averaging the signal, are the free parameters in this analysis. This section encompasses the details of the implementation and effectiveness of the formalism. Section \ref{sec:time delay} briefs about determining the delay time, while Section \ref{sec:delta t} deals with calculating the optimum averaging time. Application on glitches in data is shown in Section \ref{sec:glitch}, and the minimum sample length required has been calculated in Section \ref{sec:sample length}.

\subsection{Determining the delay time $\Delta t_{\nu}$} \label{sec:time delay}

To obtain the proper correlation, the time delay ($\Delta t_{\nu}$) should be accurately predicted. Since we are dealing with GW data sampled at $16 \si{k Hz}$ and EM data sampled at high cadence, varying from one telescope mission to another, it is very computationally expensive to determine two free parameters. Consider a case with 200s of GW and EM data sampled at $16 \si{k Hz}$ and $200 \si{Hz}$, respectively. For the TDCC to work, both signals should have the same sampling frequency; one with a higher sampling frequency can always be downsampled to match the lower sampling frequency of the other signal without losing much information. After downsampling the GW data, the time complexity to perform a direct correlation with two free parameters is $\mathcal{O}(n \times m \times p)$, while the space complexity is $\mathcal{O}(n+n)$, where $n$ is the number of elements of the dataset, and $m$ and $p$ are the lengths of the parameter space of time delay and sampling time, respectively.
\begin{figure*}[t]
    \centering
    \includegraphics[width=.46\linewidth]{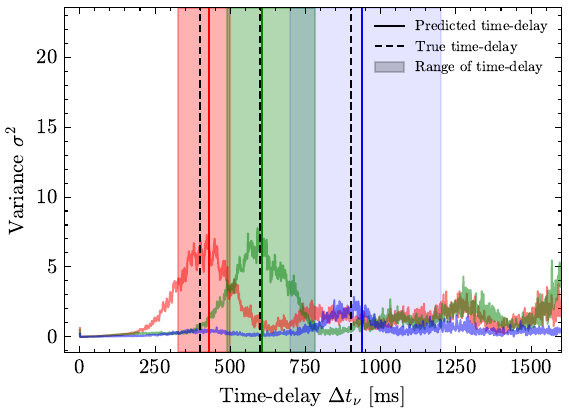}
    \qquad
    \includegraphics[width=.42\linewidth]{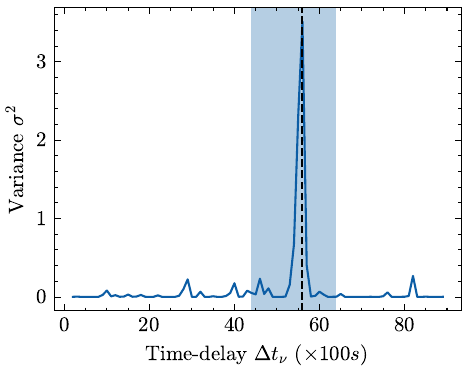}
    \caption{Left: The variance ($\sigma^2$) vs. time delay ($\Delta t_{\nu}$), for TDCC with time domain strain data (see \eqref{eqn:TDCC_ht}), for a set of three EM signals (red, green, and blue) with different time delays (0.4s, 0.6s, 0.9s, respectively). The solid lines represent the predicted delay between the GW signal and the corresponding EM counterpart, while the dashed lines correspond to the true input time delay. The shaded region corresponding to each signal (each color) gives the bound for time delays that will give out significant TDCC coefficients. Right: The variance ($\sigma^2$) vs. time delay ($\Delta t_{\nu}$), for TDCC with $\Omega_{\rm GW}(f,t)$ (see \eqref{eqn:TDCC_Omega}), for EM signal with a time delay of 5600s. The dashed black line represents the predicted delay between the GW signal and the corresponding EM counterpart.}
    \label{figp:time-delay-sgwb_deltat}
\end{figure*}

To optimize the process of estimating the time delay, the two independent variables are consolidated into a single variable by equating them, which reduces the time complexity to compute the one to $\mathcal{O}(n \times m)$. Correlations will only be seen if the time delay is accurately predicted, although there will be a substantial error when the averaging time scale is comparable with the time delay. However, once the time delay has been resolved, the only remaining variable is the sampling time, which can be easily determined with further analysis as shown in Section \ref{sec:delta t}.

After reducing the parameter space to one, we determine the time delay. The TDCC coefficients will show a peak for correct time-delay values, while they will vanish for incorrect values. We choose variance as our statistic to determine the proper time delay. We calculate the variance of the TDCC corresponding to a given $\Delta t_{\nu}$ as follows.
\begin{equation} \label{eq:variance}
    \sigma^2 = \frac{\sum\limits_{i=1}^{n} (C_{\nu}(t_{i}, \Delta t_{\nu})- \overline{C}_{\nu}(\Delta t))^2}{n-1},
\end{equation}
where $C_{\nu}(t_{i}, \Delta t_{\nu})$ is the value of the TDCC coefficient at time $t_i$, $\overline{C}_{\nu}(\Delta t)$ is the mean of $C_{\nu}(t_{i}, \Delta t_{\nu})$.

We simulate a single BNS GW strain signal with three EM counterparts with different time delays of $0.4$s, $0.6$s, and $0.9$s. We generate a GRB signal (see Section \ref{subsec:grb170817a}) with a time delay of $0.4$s, $0.6$s, and $0.9$s after the GW trigger. This EM signal is correlated with GW flux obtained from the simulated BNS strain signal with source properties $z=0.4$, $m_1 = 1.63$, and $m_2 = 1.23$. The left plot in Figure \ref{figp:time-delay-sgwb_deltat} shows how the variances ($\sigma^2$) of TDCC, with time domain strain data (see \eqref{eqn:TDCC_ht}), vary with $\Delta t_{ \nu}$. We calculate variance $\sigma^{2}$ for a range of time delays as shown in the left panel of Figure \ref{figp:time-delay-sgwb_deltat} for a correlation between flux from the GW strain signal and its EM counterparts. The distribution of variances peaks at values of correct time delay in agreement with the true time delay (dashed lines). The shaded region corresponding to each peak gives the uncertainty in determining $\Delta t_{\nu}$; however, we will always choose the peak value (solid lines) to mitigate the errors. 

For TDCC with $\Omega_{\rm GW}(f,t)$ (see \eqref{eqn:TDCC_Omega}), we perform a similar analysis. We use SGWB data, averaged over $100$s, to correlate with EM data. The corresponding analysis for determining the time-delay results is in the right panel of Figure \ref{figp:time-delay-sgwb_deltat}. However, in this case, we cannot use the averaging time $\Delta t_{\nu} < 100$s since the data is already averaged over 100s. This analysis of SGWB results, which has an EM counterpart with a delay of 5600s, also gives the maximum variance for the correct predicted time delay.

\subsection{Determining the optimum averaging time $\delta t$} \label{sec:delta t}

If the time delay is already known, the averaging time, $\delta t$, can be considered a free parameter with a reduced parameter space. If $\delta t$ is chosen to be very small, it will lead to a higher level of noise in the signal, while if it is taken to be large, the correlation factor will become small due to the presence of an uncorrelated signal. 

\begin{figure*}[t]
    \centering
    \includegraphics[width=.46\linewidth]{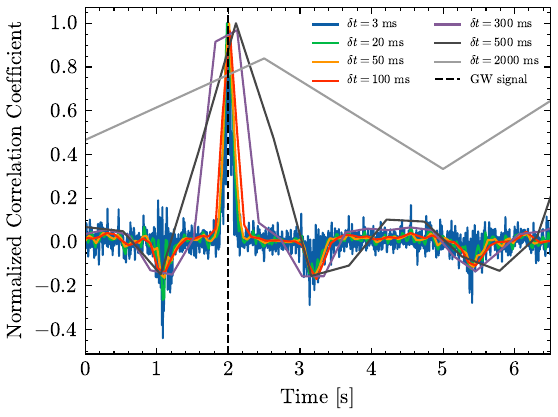}
    \qquad
    \includegraphics[width=.46\linewidth]{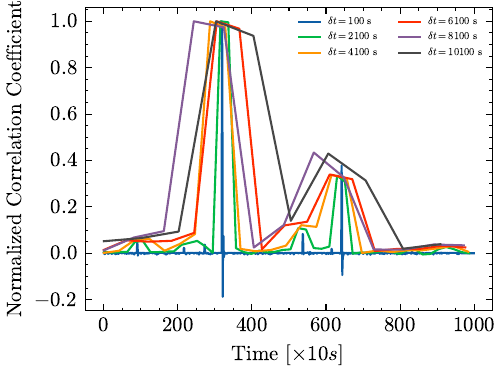}
    \caption{Left: normalized TDCC with time domain strain data (see \eqref{eqn:TDCC_ht}) vs. time in seconds for a GW signal at $t^{\text{GW}}=2$s with EM counterpart appearing 0.2s after $t^{\text{GW}}$. The noise in the correlation factor diminishes as the averaging time is increased. Right: TDCC with $\Omega_{\rm GW}(f,t)$ data (see \eqref{eqn:TDCC_Omega}) coefficient vs. time with EM counterpart appearing at 3400s after $t^{\text{GW}}$. The noise in the correlation factor diminishes as the averaging time is increased. $\delta t$ here corresponds to the averaging time.}
    \label{figp:deltat-sgwb_deltat}
\end{figure*}

We check for the effect of choosing an averaging-time scale by simulating a GW event with $m_1 = 1.87 M_\odot$, $m_2 = 1.12 M_\odot$, $z=0.24$ occurring at $t^{\text{GW}}=2$s and its EM counterpart occurring at $t^{\text{GW}} + 0.2$s. The left panel of Figure \ref{figp:deltat-sgwb_deltat} shows the normalized time-domain cross-correlation coefficient against time delay for different time delay values, $\delta t$. Choosing a $\delta t$ value below 20 ms results in sharper peaks, but it also leads to higher noise levels in the signal, as expected. It becomes evident that as the averaging time $\delta t$ increases, the noise in the correlation coefficient diminishes. However, the peak position becomes difficult to interpret if the averaging period becomes too large (greater than 500ms). Therefore, selecting an optimal $\delta t$ is crucial: a smaller $\delta t$ enhances resolution but increases noise. In comparison, a larger $\delta t$ reduces noise but at the cost of potentially smoothing out important signal features. Similar results, see the right panel of Figure \ref{figp:deltat-sgwb_deltat}, are obtained when we repeat the analysis for an SGWB signal that has an EM counterpart 3400s after the trigger. Such an analysis is not necessary for the SGWB signal where the time delay for the EM signal is less than the integration time scale, and a point-to-point ($\delta t = 1$) correlation will be required to get accurate results.

An optimum value for the averaging time can be obtained when we get distinct peaks in the data, i.e., with maximum points lying at zero. We can constrain the parameter space of values for averaging time by making the requirement of averaging time smaller than the time delay between the GW and EM signal, but not too small so that noise becomes dominant. 

\subsection{Spurious Glitches in the data} \label{sec:glitch}

\begin{figure}
    \centering
    \includegraphics[width=\linewidth]{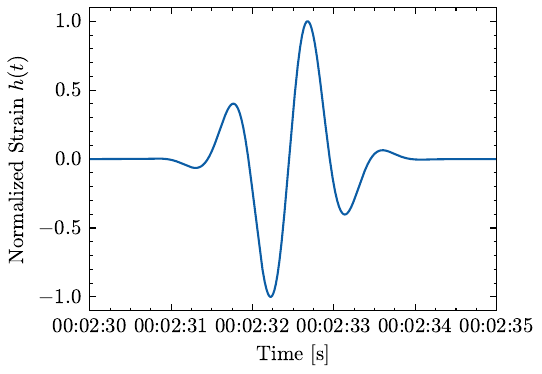}
    \caption{We show a simulated waveform of a blip glitch as a normalized strain $h(t)$ as a function of time.}
    \label{figp:spurious_glitch}
\end{figure}

\begin{figure*}[t]
    \centering
    \includegraphics[width=\textwidth]{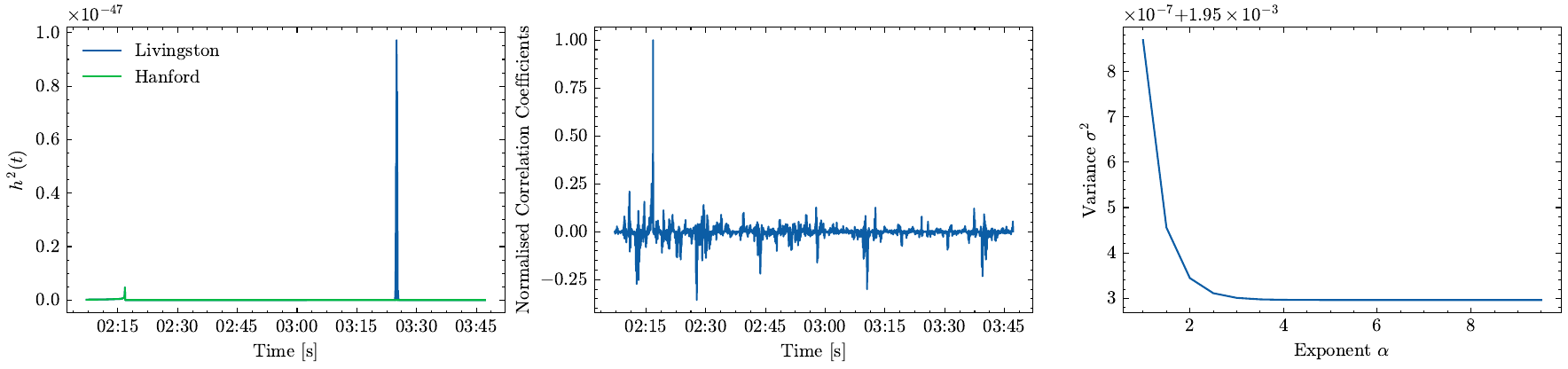}
    \caption{Left: $h^2(t)$ vs. time [s] plot for simulated data from the Hanford and Livingston observatories with a blip glitch present in the data from Livingston. It is assumed that there is no time delay between the GW trigger between both detectors. Center: Normalized correlation coefficient vs. time [ms] plot for the Hanford and Livingston observatory data. No glitch is observed in the correlated data. Right: Variance $\sigma^2$ ($\times 10^{-4}$) vs. Exponent $\alpha$ plot to check the effectiveness of the formalism for blip glitches with a varying SNR. A higher value of variance means the glitch dominates the correlation coefficients.}
    \label{figp:spurious_glitch2}
\end{figure*}

A noise transient is classified as a blip glitch if it is a very short-duration transient \citep{Abbott_2016_glitch,Abbott_2018_glithc}, $\mathcal{O}(10)$ ms, with a large frequency bandwidth, $\mathcal{O}(100)$ Hz. GW detectors are prone to blip glitches \citep{Sorazu_2010} occurring due to seismic, acoustic, and electromagnetic disturbances near the interferometers or noise arising from the detector controls \cite{cabero_blip_2019}. The characteristic time-domain shape of a blip glitch can resemble the GW signal (see Figure \ref{figp:spurious_glitch}). While blip glitches are uncorrelated between two detectors, they contribute to the background of transient GW. In this section, we show how formalism can be used to rule out spurious glitches in the data.

Consider two detectors, Hanford and Livingston, with a blip glitch occurring in the Livingston detector sometime after a GW signal is detected (see the left panel in Figure \ref{figp:spurious_glitch2}). We can perform the TDCC between data from Livingston's signal containing a glitch and Hanford's signal. It is assumed that the time delay between the GW signals detected at the two different detectors is zero. 

Applying the formalism to GW data from two detectors with the known time delay $t_d^{\text{phy}}$ outputs the resultant coefficients, which are devoid of the glitch and have a strong correlation only at the trigger of the GW signal (see the center panel in Figure \ref{figp:spurious_glitch2}). Since the glitches in two different detectors are uncorrelated, while a GW signal is, the glitch should not show up in the cross-correlation, and this is exactly what is happening here. A small residue is obtained, however, at the position of the glitch, which is a result of noise. The glitch here is roughly $10^{-2}$ times the GW signal; however, there may be times when the glitch is even stronger than the signal. For a strongly correlated signal, the variance of the TDCC coefficients is minimal ($\sigma^2 \sim 10^{-5}$). Let the strength of the glitch be 1; note that the $h^2(t)$ is of the order of $10^{-48}$.

To see the effects of varying glitch strength, we multiply it by a variable factor $10^{-\alpha}$. The presence of a glitch in TDCC will result in additional peaks. Such peaks will result in higher variance statistics of the coefficients. We calculate such variances for different values of $\alpha$. A higher variance value means the glitch dominates the correlation coefficients. From the rightmost panel in Figure \ref{figp:spurious_glitch2}, the order of variances obtained is very small. Thus, we can practically mitigate all the glitches, given that we know the physical time delay. Hence, this formalism can also extract `glitch-less' data by correlating data from two or more detectors, which in turn can be used to correlate with EM signals.

\subsection{Minimum sample length required for TDCC} \label{sec:sample length}

\begin{figure*}
    \centering
    \includegraphics[width=\textwidth]{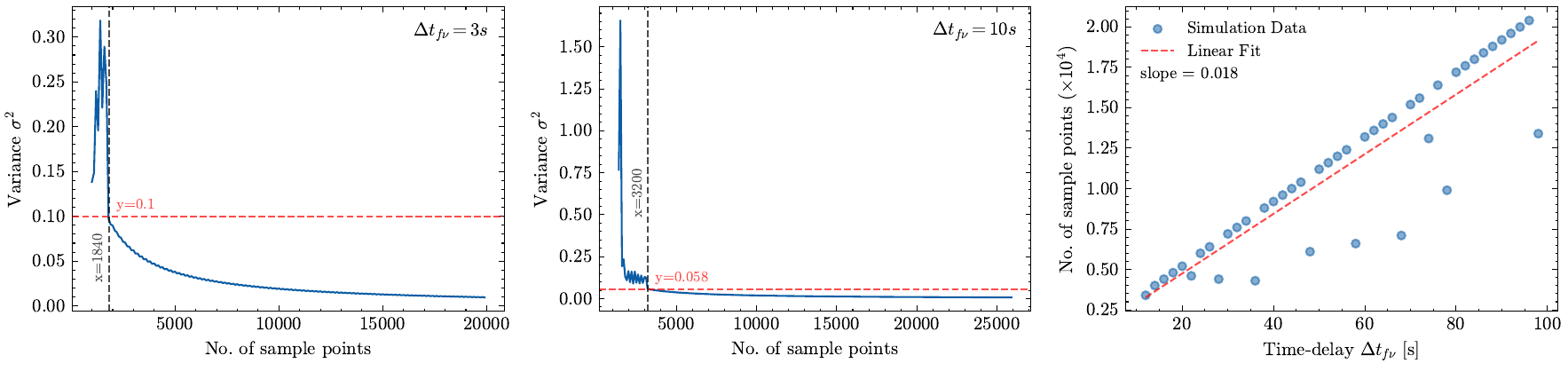}
    \caption{Left: Variance $\sigma^2$ vs. number of sample points for a GRB signal with $\Delta t_{\nu}=3$s. The variance drops at $x=1840$. Center: Variance $\sigma^2$ vs. Number of sample points for a GRB signal with $\Delta t_{\nu}=10$s. The variance drops at $x=3200$. Right: Number of sample points ($\times 10^{4}$) wrt time delay $\Delta t_{\nu}$ [s] between GW and GRB signal. The scattered points show the point after which the variances show a decline for corresponding $\Delta t_{\nu}$. The observed slope is 0.018.}
    \label{figp:threshold}
\end{figure*}

The TDCC approach heavily relies on the notion of averaging. We cannot deny the possibility that sometimes the signals coming from observations can be short in length. Cross-correlating such signals, which will involve averaging, may result in errors in results and wrong correlations. It is necessary to know what minimum sample length of the signal is required so that we do not lose any information while averaging and the results are meaningful. Consider the scenario with a GW signal, and a GRB signal with a delay of $3$s after the GW trigger. We examine the variance of the normalized TDCC coefficient against the number of sample points, i.e., the length of the signal, as can be seen in the left panel of Figure \ref{figp:threshold}. At around 1840 sample points, there is a sharp peak in the normalized TDCC coefficient variances, followed by a gradual exponential-like decline. The higher value of variances implies a significant deviation from zero, meaning there is more than one peak for a singly correlated signal, which hints toward incorrect correlation. Given a sample frequency of $100 \si{Hz}$, the corresponding time interval is 18.4 seconds. We achieve comparable results by increasing the time delay between the signals and performing a similar analysis. The minimum sample length for a signal with a temporal delay of $10$s is $32$s (See the center panel of Figure \ref{figp:threshold}). A general trend can be observed by extending the analysis to an array of time-delays. The right-most panel in Figure \ref{figp:threshold} shows a general trend with the scattered points showing the point after which the variances show a decline. A few points show deviation from the general trend; however, these points indicate that they converge to a lower variance sooner, although taking more sample points would not affect the results. Hence, we require a data stream that is twice as long as the expected time delay to obtain meaningful results.

\section{Analysis Technique for SGWB} 
\label{apx:sgwb}

This section gives a brief overview of standard analysis techniques for SGWB data. The data in the time domain from detector $I$ can be written as
\begin{equation}
    s_I(t) = h_I(t) + n_I(t),
\end{equation}
where $h_I(t)$ and $n_I(t)$ are the GW strain and detector noise, respectively. The detector's response changes as its antenna pattern sweeps across the anisotropic distribution of gravitational radiation. To account for this change, we split the data into small chunks of duration $\tau$ such that the detector response function does not change over the interval. This $\tau$ should be much greater than the light travel time between any pair of detectors. For LIGO, $\tau$ can be chosen from $100s$ to $1000s$. The short-time Fourier transform (STFT) of the GW signal at a particular time $t$ can be written as
\begin{equation}
    \tilde{s_{I}}(t,f) = \int_{t-\tau/2}^{t+\tau/2} \mathrm{d} t'~ s_{I}(t') e^{-i 2\pi f t'}.
\end{equation}

The data from two different detectors is cross-correlated to map the GW strain from the noise-covered signal. Since the noise in the detectors is expected to be uncorrelated, i.e.,
\begin{equation}
    \langle n_I(t,f)n_J^*(t,f) \rangle = 0,
\end{equation}
where, $I$ and $J$ are two different detectors. The expectation of the cross-correlation can thus be written as
{\begin{equation}
    \langle \Omega_{IJ}(t,f) \rangle \equiv \frac{2}{\tau} \langle s_I(t,f) s_J^*(t,f) \rangle = \frac{2}{\tau} \langle h_I(t,f)h_J^*(t,f) \rangle.
\end{equation}}

\begin{figure*}
    \centering
    \includegraphics[width=\textwidth]{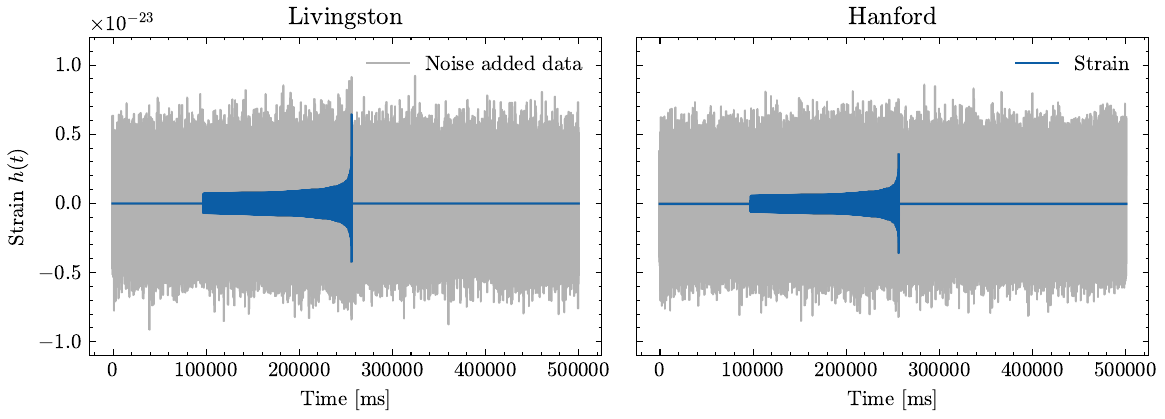}
    \caption{Simulated BNS merger GW strain $h(t)$ vs. time [ms] for LIGO-Livingston (left) and LIGO-Hanford (right). The GW signal is completely absorbed in the detector noise; this signal can be classified as SGWB.}
    \label{figp:sgwb_strain}
\end{figure*}

\begin{figure*}
    \centering
    \includegraphics[width=\textwidth]{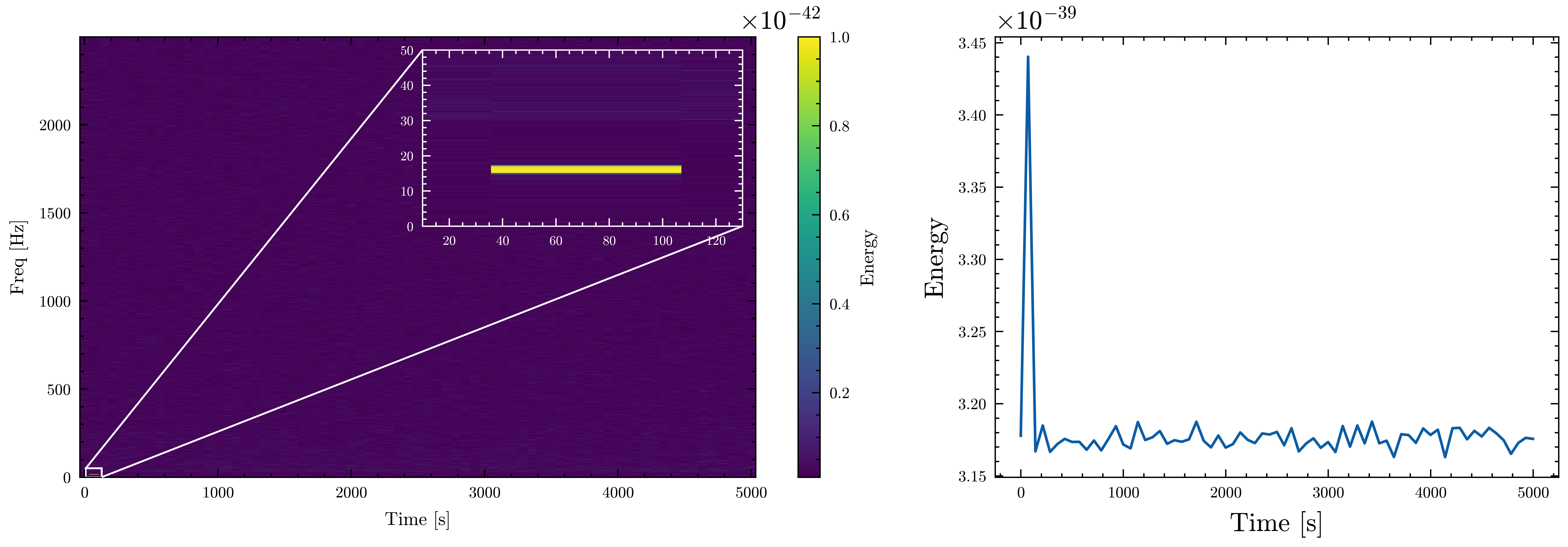}
    \caption{Left: Spectrogram for SGWB extracted using short-time Fourier transform. Right: Energy vs. time [s] obtained from the spectrogram for the SGWB signal. This extracted signal from the noise can cross-correlate with its EM counterparts.}
    \label{figp:sgwb_signal}
\end{figure*}

Consider a BNS merger at a redshift $z=0.2$; such events are blind to the LVK detectors and are shrouded in the GW noise, as shown in Figure \ref{figp:sgwb_strain}. The spectrogram (see the left panel in Figure \ref{figp:sgwb_signal}) results from the above analysis technique. A short yellow bar masked in the vast background of noise indicates the presence of a signal detected at both detectors. The energy of this GW signal ($\Omega_{\text{GW}}$) is plotted in the right panel of Figure \ref{figp:sgwb_signal}. This extracted signal can cross-correlate with its detected EM counterparts to bring up any possible association between them.

\bibliography{prd}

\end{document}